\documentclass[lettersize,journal]{IEEEtran}

\IEEEoverridecommandlockouts
\usepackage{amsthm,amsmath,amssymb}
\usepackage{amsfonts}
\usepackage{caption}
\usepackage{graphicx}
\usepackage{stfloats}
\usepackage[font=small]{caption}
\usepackage{cite}
\usepackage{booktabs}
\usepackage[labelformat=simple]{subcaption}
\usepackage{soul}
\usepackage{tabularx}
\usepackage{textcomp}
\usepackage{mathtools}
\usepackage{color}
\usepackage[dvipsnames]{xcolor}
\usepackage[normalem]{ulem}
\usepackage[thicklines]{cancel}

\newcommand{\mbf}{\mathbf}
\newcommand{\mrm}{\mathrm}
\newcommand{\bsm}{\boldsymbol}
\newcommand{\mal}{\mathcal}
\newcommand{\mbb}{\mathbb}

\DeclareMathOperator*{\argmax}{argmax}
\usepackage{array}
\usepackage{algorithmic}
\usepackage{amsmath}

\usepackage{graphicx,float}

\usepackage[utf8]{inputenc}
\usepackage{enumerate}
\usepackage{color}
\usepackage{amsfonts}
\usepackage{algorithm}
\usepackage{algorithmic}

\def\BibTeX{{\rm B\kern-.05em{\sc i\kern-.025em b}\kern-.08em
    T\kern-.1667em\lower.7ex\hbox{E}\kern-.125emX}}

\begin{document}
%
\title{Dynamic Channel Knowledge Map Construction in MIMO-OFDM Systems}

%
%



\author{Wenjun Jiang,~\IEEEmembership{Graduate Student Member,~IEEE}, 
	Xiaojun Yuan,~\IEEEmembership{Fellow,~IEEE}, Chenchen Liu,~\IEEEmembership{Graduate Student Member,~IEEE}, and Boyu Teng,~\IEEEmembership{Graduate Student Member,~IEEE}}
\maketitle

%
%
%

\begin{abstract}
Channel knowledge map (CKM) is a promising paradigm for environment-aware communications by establishing a deterministic mapping between physical locations and channel parameters. Existing CKM construction methods focus on quasi-static propagation environment. This paper develops a dynamic CKM construction method for multiple-input multiple-output orthogonal frequency division multiplexing (MIMO-OFDM) systems. We establish a dynamic channel model that captures the coexistence of quasi-static and dynamic scatterers, as well as the impacts of antenna rotation and synchronization errors. Based on this model, we formulate the problem of dynamic CKM construction within a Bayesian inference framework and design a two-stage approximate Bayesian inference algorithm. In stage I, a high-performance algorithm is developed to jointly infer quasi-static channel parameters and calibrate synchronization errors from historical measurements. In stage II, by leveraging the quasi-static parameters as informative priors, a low-complexity algorithm is designed to estimate dynamic parameters from limited real-time measurements. Simulation results validate the superiority of the proposed method and demonstrate its effectiveness in enabling low-overhead, high-performance channel estimation in dynamic environments.
\end{abstract}


\section{Introduction}

The sixth-generation (6G) wireless systems is characterized by transformative trends, including the deployment of massive multiple-input multiple-output (MIMO) arrays with hundreds to thousands of antennas, the expansion of transmission bandwidths up to $100 ~ \text{GHz}$, and terminal access densities approaching $10^8 / \text{km}^2$ \cite{sector2023framework}. These advances are envisioned to support emerging services such as live streaming, immersive virtual reality (VR) , and digital twin. On the other hand, this massive scaling significantly broadens the wireless channel's frequency and spatial dimensions, bringing several fundamental challenges to the communications physical layer (PHY). The challenges include excessive pilot overhead for channel state information (CSI) acquisition, rapidly increasing complexity of channel estimation, and expensive beam-training costs. To overcome these bottlenecks, the concept of the channel knowledge map (CKM) has recently emerged as a promising paradigm for environment-aware communications \cite{CKM_Zeng,DCT_Jiangwen}.

CKM can be regarded as a digital twin of wireless channel, capturing the deterministic mapping between spatial information (of transceiver) and the corresponding channel characteristics—such as multipath delays, angles, and gains. This scene-specific characterization fundamentally moves beyond traditional statistical channel models, reducing the randomness of wireless channel. Leveraging the scene-specific channel parameters, CKM empowers the design of  PHY algorithms. In pilot configuration, CKM provides the information of multipath richness. Dense-scattering areas can thus be allocated with more pilot resources, whereas LoS-dominated regions can be assigned fewer pilots. In channel estimation, CKM can provide  power-delay profile (PDP) and power-angle profile (PAP) to enhance estimation accuracy. In beam training, CKM can support beam index map  to reduce beam-searching complexity and overhead \cite{BF_CKM}. In interference suppression, CKM can provide the channel matrix subspace of access users and enable interference characteristics to be efficiently captured \cite{Wenjun_CKM}.

To fully harvest the potential of CKM-assisted applications, the key prerequisite lies in the construction of a high-precision CKM. The theoretical foundation for CKM construction is from electromagnetic field theory. Given the boundary conditions of a propagation environment (e.g., the geometry and material properties of scatterers), Maxwell's equations determine the channel response between any transceiver locations. Ray-tracing (RT) methods \cite{Ray} serve as classical approximations by simulating geometric optics propagation. Although RT has high accuracy, it depends on complete 3D environmental models and electromagnetic properties of materials, limiting the application to CKM-assisted PHY algorithm. To address these issues, learning-based CKM construction has attracted growing attention. Inspired by the similarity between wireless propagation and visible-light rendering, several neural rendering network were adopted to model the ray–scatterer interactions without relying on explicit electromagnetic parameters \cite{zhao_nerf2_2023,orekondy2023winert,NeRF_SS,wen_neural_2025}. Particularly, the authors in \cite{orekondy2023winert} employs neural radiance field (NeRF) \cite{mildenhall2021nerf} to learn the interaction between rays and scatterers; the authors in \cite{wen_neural_2025} employ 3D Gaussian splatting \cite{kerbl2023_3DGS} to predict site-specific multipath angles and delays. Besides neural rendering methods, CKM construction based on generative models have been proposed \cite{chi2024rf,fu2025generative}.

Existing literature focuses on quasi-static propagation environments, where the CKM characterizes the channel parameters associated with the quasi-static scatterers (such as buildings or ground). In practical scenarios, however, multiple factors contribute to channel dynamics. A primary factor is the dynamic scatterers which cause the birth and death of channel paths. To characterize this impact, the geometric structure of dynamic scatterers can be sensed and mapped to the corresponding channel parameters \cite{XiuCheng_Diff,ChengXiang_Dynamic}. Despite its effectiveness, this approach typically incurs substantial sensing overhead and high computational complexity. Beyond dynamic scatterers, high channel dynamics are also introduced by other factors, such as terminal orientation changes and system synchronization errors. As a result, constructing a dynamic CKM under limited resources remains an open and challenging problem.

This paper aims to develop an efficient dynamic CKM construction method in multiple-input multiple-output orthogonal frequency division multiplexing (MIMO-OFDM) systems, considering multiple dynamic factors. We model the dynamic channel and propose a two-stage CKM construction framework based on both historical and real-time channel observations. To ensure accuracy and efficiency, the CKM construction problem is formulated into a Bayesian inference framework, and an approximate Bayesian inference algorithm is developed to infer channel parameters. The constructed CKM is further applied to enable low-overhead, high-performance channel estimation in dynamic wireless environments. The main contributions are summarized as follows:

\begin{itemize}
	\item We establish a dynamic channel model in MIMO-OFDM systems. This model accounts for multiple dynamic factors: the coexistence of static and dynamic scatterers, the power spectral variation introduced by antenna orientation rotation, and the group delay caused by synchronization error. By integrating these factors, we derive a baseband channel model comprising quasi-static and dynamic channel parameters. 
	\item We develop a two-stage framework for constructing dynamic CKM. Stage I extracts quasi-static channel parameters from historical measurements. The BS coverage area is discretized into spatial grids, and the observations from each grid are used to extract a representative set of quasi-static channel parameters. This stage is designed to be robust against synchronization errors and power spectral variations. Stage II extracts dynamic channel parameters from real-time measurements. It leverages the quasi-static parameters obtained in stage I as informative priors, enabling low-complexity estimation with limited pilot overhead.
	\item We formulate the problem of dynamic CKM construction into a sequential Bayesian inference problem and design an approximate Bayesian inference algorithm. For stage I, we propose a joint inference of group delay and path complex coefficients, and establish a path generation mechanism to improve estimation accuracy. For stage II, we propose a low-complexity group delay calibration method based on the prior knowledge from the quasi-static parameter estimation, and design a path-wise estimation of dynamic path coefficients exploiting a sparse probability model.
\end{itemize}
Extensive simulations are conducted to validate the superiority of proposed method against several benchmarks, including the CKM construction without dynamic component estimation, without synchronization error calibration, and without quasi-static parameter priors. Particularly, it is shown that the proposed method achieves accurate channel estimation with limited pilot overhead in dynamic environments.


\emph{Organization:} In Sec. II, we introduce the dynamic channel model. In Sec. III, we establish the framework of dynamic CKM construction. In Sec. IV, we formulate the CKM construction problem into a Bayesian inference problem. In Sec. V and Sec. VI, we develop the approximate Bayesian inference algorithm for stage I and II of the CKM construction, respectively.  In Sec. VII, we provide numerical results, and in Sec. VIII, we conclude this paper.

\emph{Notation:} We use bold capital letters (e.g., $\mathbf X$) for matrices and bold lowercase letters (e.g., $\mathbf x$) for vectors. $(\cdot)^T$, $(\cdot)^*$, and $(\cdot)^H$ denote the transpose, the conjugate, and the conjugate transpose, respectively. $\delta(\cdot)$ denotes the Dirac delta function. ${\rm diag}(\mathbf x)$ denotes the diagonal matrix with the $i$-th diagonal entry being the $i$-th entry of $\mbf x$. $\mrm{tr}(\cdot)$ and $\mrm{vec}(\cdot)$ denote the trace and vectorization operators, respectively. $(\cdot)^{-1}$ denotes the matrix inverse. $\otimes$ and $\odot$ denote the Kronecker and Hadamard product, respectively. Matrix $\mbf I$ denotes an identity matrix with an appropriate size. For a random variable $x$, its pdf is denoted by $p(x)$. $\mathbb E[\cdot]$ denotes the expectation operator. The pdf of a complex Gaussian random vector $\mathbf x \in \mathbb{C}^N$ with mean $\bsm{\mu}$ and covariance $\bsm{\Sigma}$ is denoted by $\mathcal{CN}(\mathbf x ; \bsm{\mu} , \bsm{\Sigma}) = |\bsm{\Sigma}|^{-1} {\rm exp}(-(\mathbf x - \bsm{\mu})^H (\bsm{\Sigma})^{-1} (\mathbf x - \bsm{\mu}))/\pi^N$. The pdf of a von Mises (V-M) random variable $ x \in [0,2\pi)$ with mean direction $\mu$ and concentration $\kappa$ is denoted by $ \mal{VM}(x;\mu,\kappa)\!=\!\frac{1}{(2 \pi I_0(\kappa))}\exp(\kappa \cos(x-\mu))$, where $I_0(\cdot)$ is the modified Bessel function of the first kind and order $0$.

\vspace{-0.3 cm}

\section{Dynamic Channel Modeling}
\subsection{Propagation Channel Model in a Dynamic Environment} \label{sec:dynamic_scatterer}

Consider an uplink MIMO-OFDM system. The base station (BS) is equipped with a uniform rectangular array (URA) of $M = M_1 \times M_2$ antennas with half-wavelength spacing. A single-antenna user communicates with the BS over $N$ subcarriers, with a subcarrier spacing of $\Delta_f$. 

\begin{figure}[htbp]
    \centering
    \includegraphics[width=0.5\textwidth]{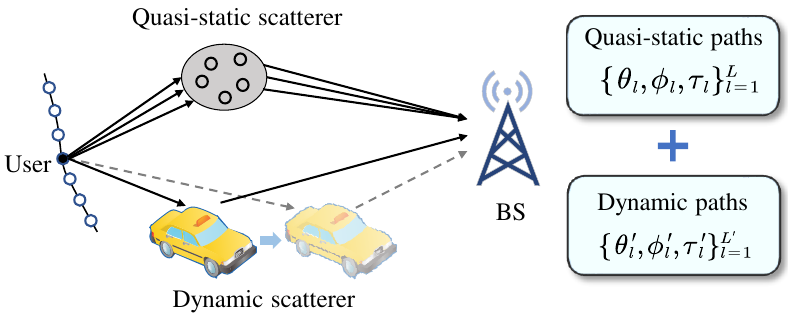}
    \caption{An example of dynamic propagation environment.}
    \label{fig:dynamic_propagation}
\end{figure}

The propagation environment consists of both quasi-static scatterers (e.g., buildings, ground) and dynamic scatterers (e.g., vehicles), as illustrated in Fig. \ref{fig:dynamic_propagation}. Correspondingly, the propagation channel is modeled as the superposition of a quasi-static channel component due to quasi-static scatterers and a dynamic channel component due to dynamic scatterers.

\vspace{-0.2 cm}
\vspace{-0.2 cm}

For the quasi-static channel component, we consider a geometry-based multi-path channel model. Specifically, we denote the parameters of the $l$-th path by its delay $\tilde{\tau}_{l}$, azimuth angle-of-arrival (AOA) $\tilde{\theta}_{l}$, and zenith AOA $\tilde{\phi}_{l}$. These are normalized as $\tau_{l} \triangleq 2 \pi \Delta_f \tilde{\tau}_{l}$, ${\theta}_{l} \triangleq \pi \sin \tilde{\theta}_{l} \cos \tilde{\phi}_{l} $, and $\phi_{l} \triangleq \pi \sin \tilde{\phi}_{l}$. We define the user's position as $\mbf p \in \mathbb{R}^3$. Since the BS is typically fixed after deployment, we can construct a mapping:
\begin{align}
 \mal{M}(\cdot) : \mbf p  \rightarrow \{ \tau_l(\mbf p), \theta_l(\mbf p), \phi_l(\mbf p)\}_{l=1}^{L}, \label{p_map}
\end{align}
where $L$ is the number of quasi-static paths. Strictly speaking, the mapping $\mal{M}(\cdot)$ varies over time due to the change of the quasi-static scatterers, e.g., building deformation or climate variations. This variation typically occurs from minutes to days. For notational simplicity, we omit the time index of $\mal{M}(\cdot)$ and refer to $\{\tau_l(\mbf p), \theta_l(\mbf p), \phi_l(\mbf p)\}_{l=1}^{L}$ as the \emph{quasi-static channel parameters}.

\begin{figure}[htbp]
    \centering
    \includegraphics[width=0.5\textwidth]{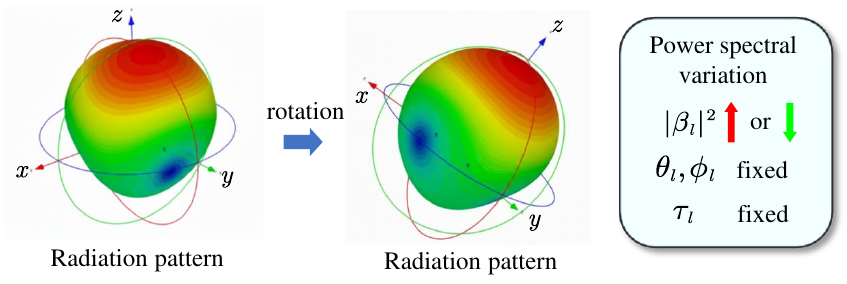}
    \caption{Illustration of power spectral variation due to antenna rotation.}
    \label{fig:antenna_rotation}
        \vspace{-0.2 cm}
\end{figure}

To characterize the complex path coefficients of quasi-static channel component, we account for the user's antenna orientation. Denote by $\Delta\psi_{h}$ and $\Delta\psi_{v}$ the normalized horizontal and vertical rotation angles, respectively. The user's antenna is taken as the origin of the reference coordinate system. For given zenith angle $\theta$ and azimuth angle $\phi$, we introduce the vector field radiation pattern as $ \mathbf{F}(\theta, \phi) = (F^{[v]}(\theta, \phi), F^{[h]}(\theta, \phi)  )^T$
where $F^{[v]}$ and $F^{[h]}$ are the complex responses for vertical and horizontal polarizations, respectively. Let $\mathbf{F}_{\mrm t}$ and $\mathbf{F}_{\mrm r}$ be the radiation patterns of the user and BS antennas, respectively. For the $l$-th path, let $( \theta_l^d(\mbf p), \phi_l^d(\mbf p) )$ be the angle-of-departure (AoD) from the user. Define the user pose as $\bsm{\chi}=[ \mbf p^T, \Delta\psi_{h}, \Delta\psi_{v}]^T$. Then, the $l$-th channel coefficient is modeled as \cite{TR38901}
\begin{align}
        \beta_l(\bsm{\chi}) = & \sqrt{P_l} \cdot \mathbf{F}_{\mrm r}^T(\theta_l(\mbf p),\phi_l(\mbf p)) \mathbf{M}_l \notag \\
        & \times  \mathbf{F}_{\mrm t}( \theta_l^d(\mbf p)+\Delta\psi_{h},\phi_l^d(\mbf p)+\Delta\psi_{v} ) \cdot e^{-j \frac{2\pi}{\lambda} d_l}, \label{eq:path_gain_detailed}
\end{align}
where $\lambda$ is the carrier wavelength, $P_l$ is the path-loss (excluding antenna gains), $d_l$ is the path length, and $\mathbf{M}_l$ is the $2 \times 2$ polarization coupling matrix for the $l$-th path. From \eqref{eq:path_gain_detailed}, an antenna rotation alters the transmit radiation pattern $\mathbf{F}_{\mrm t}( \cdot,\cdot)$ for each departure angle. This, in turn, causes variation in the path power $|\beta_l(\bsm{\chi})|^2$, as illustrated in Fig. \ref{fig:antenna_rotation}.

\vspace{-0.2 cm}

With the geometric parameters from \eqref{p_map} and the complex coefficients from \eqref{eq:path_gain_detailed}, the quasi-static channel is expressed as
\begin{align}
    \mbf{H}^{\mrm{s}}(\bsm{\chi}) = \sum_{l=1}^{L} \beta_l(\bsm{\chi}) \mbf a_N(\tau_{l}(\mbf p)) \left(  \mbf a_{M_1}(\theta_{l}(\mbf p)) \otimes \mbf a_{M_2}(\phi_{l}(\mbf p)) \right)^T, \label{Hprop}
\end{align}
where $\mbf a_{x}(\omega)$ is the steering vector defined as
\begin{align}
    \mbf a_{x}(\omega) \triangleq  \frac{1}{\sqrt{x}} [ 1, e^{-j \omega }, ... , e^{-j  (x-1)  \omega } ]^T. \notag
\end{align}

For the dynamic channel component, if the geometry information (e.g., pose, surface profile) of dynamic scatterers are available, one can construct a mapping from the user's position and the scatterers' geometry information to the dynamic channel parameters. However, tracking such geometry information is practically difficult. 
For simplification, we model the dynamic channel component as a function of time-slot $t$:
\begin{align}
    \mbf{H}^{\mrm{d}}(t) = \sum_{l=1}^{L'} \beta'_{l}(t)\, \mbf a_N(\tau'_{l}(t)) \left( \mbf a_{M_1}(\theta'_{l}(t)) \otimes \mbf a_{M_2}(\phi'_{l}(t)) \right)^T, \label{eq:H_dyn_single_path}
\end{align}
where $\{\tau'_{l}(t)$, $\theta'_{l}(t),\phi'_{l}(t)\}_{l=1}^{L'}$ are the \emph{dynamic channel parameters}; $L'$ is the number of potential dynamic paths; The complex coefficient $\beta'_{l}(t)$ encapsulates both the path's activity and its amplitude/phase, with $\beta'_{l}(t)=0$ indicating that the path is inactive at time-slot $t$. Compared to $\mbf{H}^{\mrm{s}}(\bsm{\chi})$ in \eqref{Hprop}, the parameters of $\mbf{H}^{\mrm{d}}(t)$ vary on a much shorter time scale, typically on the order of seconds or even milliseconds, due to the motion of the dynamic scatterers (e.g., cars).

Assume that there exists a mapping from time-slot $t$ to user's pose $\bsm{\chi}$, i.e., $f: t \rightarrow \bsm{\chi}$. With \eqref{Hprop} and \eqref{eq:H_dyn_single_path}, the overall propagation channel at time-slot $t$ is expressed as
\begin{align}
    \mathbf{H}(t) = \mathbf{H}^{\mrm{s}}(\bsm{\chi}) + \mathbf{H}^{\mrm{d}}(t).
\end{align}

\vspace{-0.2 cm}
\subsection{Baseband Channel Model under Synchronization Error} \label{sec:sync_error}



\begin{figure}[htbp]
	\centering
	\includegraphics[width=0.5\textwidth]{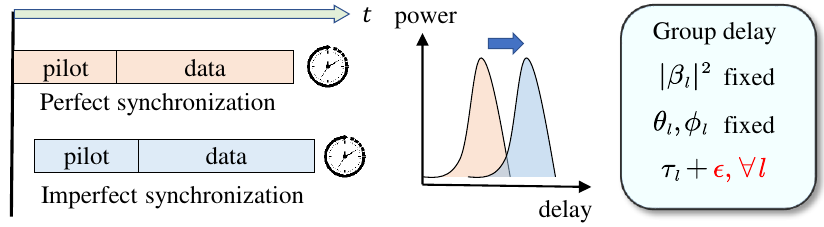}
	\caption{An illustration of synchronization error.}
	\label{fig:sync_error}
\end{figure}

Sec. II-A provides the propagation channel model considering dynamic scatterers and antenna rotation. We further consider the baseband channel model and introduce the synchronization error. Specifically, the user's local clock is not perfectly synchronized with the BS clock. This mismatch is represented by a time-varying \emph{clock offset} $\gamma^{\mathrm{offset}}(t)$\footnote{Besides clock offset, clock skew (i.e., clock-rate mismatch) also contributes to synchronization errors. Since the skew is typically close to one \cite{SLS}, its intra-slot impact is negligible, while its inter-slot cumulative effect is absorbed into the time-varying clock offset $\gamma^{\mathrm{offset}}(t)$.}. In cellular networks, timing advance (TA) is employed to compensate for the propagation delay such that uplink signals from different users arrive at the BS in a synchronized manner \cite{TS38533}. Specifically, the BS measures the uplink arrival time and commands the user to adjust its transmission timing, where a user farther from the BS is instructed to transmit earlier (i.e., with a larger TA) to offset the longer propagation delay. With the clock offset and TA, the synchronization error is given by
\begin{align}
    \tilde{\epsilon}(t) = \tau_1 - \gamma^{\mathrm{TA}}(t) - \gamma^{\mathrm{offset}}(t).
\end{align}
where $\gamma^{\mathrm{TA}}(t)$ is the TA, and $\tau_1$ is the propagation delay of the first path. Without loss of generality, assume $\tau_1$ is the smallest path delay in $\{\tau_l\}_{l=1}^L$. Perfect synchronization, i.e., $\tilde{\epsilon}(t)=0$, is achieved when $\gamma^{\mathrm{offset}}(t)=0$ and $\gamma^{\mathrm{TA}}(t) = \tau_1$. In practice, imperfections in TA estimation and clock drift lead to a non-zero $\tilde{\epsilon}(t)$, which can be on the order of several hundred nanoseconds. This error introduces a group delay to all propagation paths, as illustrated in Fig. \ref{fig:sync_error}.

We now integrate the effects of antenna rotation, dynamic scatterers, and synchronization error into an equivalent baseband channel model. Define the normalized group delay offset as $\epsilon(t) \triangleq 2 \pi \Delta_f \tilde{\epsilon}(t)$. With $\epsilon(t)$, \eqref{Hprop}, and \eqref{eq:H_dyn_single_path}, the baseband channel model is expressed as
\begin{align}
	 \mbf{H}^{\mrm{BB}}(t) 
	= & \underbrace{\sum_{l=1}^{L} \beta_l(\bsm{\chi}) \mbf a_N(\tau_l(\mbf p) + \epsilon(t)) \mbf b(\theta_l(\mbf p),\phi_l(\mbf p))^T}_{\mbf{H}^{\mrm{BB},\mrm{s}}(t)} + \notag \\
	& \underbrace{\sum_{l=1}^{L'} \beta'_{l}(t) \mbf a_N(\tau'_{l}(t) + \epsilon(t)) \mbf b(\theta'_{l}(t),\phi'_{l}(t))^T}_{\mbf{H}^{\mrm{BB},\mrm{d}}(t)}, \label{eq:H_BB}
\end{align}
where $\mbf b(\theta,\phi) \triangleq \mbf a_{M_1}(\theta) \otimes \mbf a_{M_2}(\phi)$. In this overall model, the first summation represents the contribution from the quasi-static scatterers, where the path coefficient $\beta_l$ incorporate the effects of antenna rotation. The second summation captures the transient effects of dynamic scatterers. Critically, the normalized group delay $\epsilon(t)$ is added to the delays of all paths within the frequency-domain steering vector $\mbf a_N(\cdot)$. This baseband model provides a complete representation that serves as the foundation for the subsequent CKM construction.

\section{Dynamic CKM Construction}

\subsection{Two-Stage Construction Framework}

Recognizing the disparity in the time scales of the changes of quasi-static and dynamic parameters, we propose a two-stage framework for the dynamic CKM construction.

\begin{figure}[htbp]
	\centering
	\includegraphics[width=0.5\textwidth]{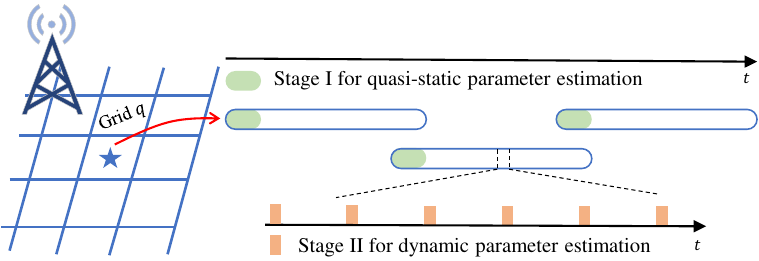}
	\caption{An illustration of two-stage CKM construction framework.}
	\label{CKM_Framework}
	\vspace{-0.2 cm}
\end{figure}

\textbf{Stage I: Quasi-Static Parameter Estimation.} This stage operates on a time scale of minutes to days. The objective is to construct a map from $\mbf p$ to $\{\tau_l(\mbf p),\theta_l(\mbf p),\phi_l(\mbf p)\}_{l=1}^L$, which can leverage large volumes of historical measurement data collected at the BS. Note that it is infeasible to directly construct the continuous mapping $\mal{M}(\cdot)$ over the entire $5$-dimensional continuous space (3D position + 2D orientation). To address this, we partition the BS coverage area into discrete grids. Within each grid, we estimate a single set of representative channel parameters by aggregating historical data from various poses in this grid. This results in a grid-level mapping of quasi-static parameters, which is formally described in Sec. \ref{sec:CKMA}.

\textbf{Stage II: Dynamic Parameter Estimation.} This stage operates on a much shorter time scale, from milliseconds to seconds. It aims to update the complex coefficients $\beta_l(\bsm{\chi})$ and $\beta_l'(t)$, estimate the dynamic parameters $\{\tau'_{l}(t), \theta'_{l}(t), \phi'_{l}(t)\}_{l'=1}^{L'}$, and calibrate the group delay $\epsilon(t)$. This process relies on periodic sounding signals (e.g., demodulation reference signal (DM-RS) or sounding reference signal (SRS) \cite{TS38211}) and should be performed with low complexity and limited pilot overhead. Compared to the grid-level parameter extraction in stage I, stage II performs user-specific parameter extraction. It leverages the estimation of quasi-static parameters (in stage I) as a strong informative prior. A detailed comparison between the two stages is provided in Table \ref{tab:static_dynamic_comparison}.

\begin{table*}[htbp]
	\centering
	\caption{Comparison between Quasi-Static and Dynamic Parameter Estimations}
	\label{tab:static_dynamic_comparison}
	\begin{tabularx}{\textwidth}{>{\raggedright\arraybackslash}p{0.25\textwidth} >{\raggedright\arraybackslash}X >{\raggedright\arraybackslash}X}
		\toprule
		\textbf{Feature} & \textbf{Quasi-Static Parameter Estimation} & \textbf{Dynamic Parameter Estimation} \\
		\midrule
		\textbf{Concept} & Grid-level. & User-specific. \\
		\textbf{Prior Knowledge} & Typically assume non-informative priors. & Leverage the estimates of quasi-static parameters as informative priors. \\
		\textbf{Update Frequency} & Update infrequently (e.g. minutes to days) due to quasi-static environmental changes. & Update frequently (e.g., milliseconds to seconds) due to dynamic scatterers, antenna rotations, and synchronization errors. \\
		\textbf{Measurement Data Requirement} & Leverages historical datasets across various terminals and poses per grid. & Relies on real-time measurements with limited pilot overhead. \\
		\textbf{Computational Complexity} & Relatively high tolerance due to infrequent update. & Relatively low tolerance for real-time update. \\
		\bottomrule
	\end{tabularx}
\end{table*}
\vspace{-0.2 cm}

\subsection{Stage I: Estimation of Quasi-Static Parameters} \label{sec:CKMA}

In this subsection, we aim to establish a grid-level mapping of quasi-static parameters. Specifically, we divide the BS coverage into $Q$ discrete grids, where each grid corresponds to a region $\mal{R}_q$ for $q=1,2,...,Q$, such that $\cup_{q=1}^Q \mal{R}_q$ covers the entire coverage area. For each grid $\mal{R}_q$, we map all poses within it to a set of channel parameters $\{\tau_l^{\mrm{s}},\theta_l^{\mrm{s}},\phi_l^{\mrm{s}}\}_{l=1}^{L^{\mrm{s}}}$, where $L^{\mrm{s}}$ is the number of grid-level quasi-static parameters. This establishes a grid-level mapping as
\begin{align}
 \mal{C}(\cdot) : q \rightarrow \{ \tau_l^{\mrm{s}},\theta_l^{\mrm{s}},\phi_l^{\mrm{s}} \}_{l=1}^{L^{\mrm{s}}}. \label{CKM}
\end{align}
Compared to \eqref{p_map}, $\mal{C}(\cdot)$ is a mapping from the grid index $q$ to the grid-level ``average" channel parameters. This mapping can be stored in the form of a look-up table. In CKM-assisted applications, when the BS acquires the user's grid, it can retrieve $\{ \tau_l^{\mrm{s}},\theta_l^{\mrm{s}},\phi_l^{\mrm{s}} \}_{l=1}^{L^{\mrm{s}}}$ through the look-up table.

The feasibility of mapping \eqref{CKM} is guaranteed by the principle of spatial consistency. Spatial consistency is a fundamental property of wireless channels, where parameters including path delays, angles, and powers vary smoothly with the user's pose. For example, in the 3GPP standard channel model \cite{TR38901}, a delay or angle parameter $\vartheta$ at positions $\mathbf{p}_1$ and $\mathbf{p}_2$ satisfies
\begin{equation}
\quad \mathrm{Cov}\big(\vartheta(\mathbf{p}_1), \vartheta(\mathbf{p}_2)\big) = \sigma^2 \cdot \rho\left(\|\mathbf{p}_1 - \mathbf{p}_2\|\right), \notag
\end{equation}
where \( \rho(\cdot) \) is a correlation function and \( \sigma^2 \) is the marginal variance. A typical choice is the exponential correlation
\(
\rho(d) = \exp\left(-\frac{d}{d_{\text{cor}}}\right),
\)
with \( d_{\text{cor}} \) denoting the correlation distance, which quantifies how fast the channel parameters decorrelate in space. According to \cite[Table 7.6.3.1-2]{TR38901}, the correlation distance for UMa NLoS and UMa LoS is $50$ m and $40$ m, respectively. Spatial consistency is also reflected in ray-tracing channel model:
\begin{equation}
	\mbf{H}^{\mrm{s}}(\bsm{\chi}) = \mathcal{F}_{\text{RT}}(\bsm{\chi}; \text{scene geometry, materials}). \notag
\end{equation}
$\mathcal{F}_{\text{RT}}$ is realized by simulating electromagnetic wave propagation within a 3D digital map of the environment. Since users in close pose experience similar scattering environments, their channel parameters vary smoothly as $\bsm{\chi}$ changes.

With \eqref{CKM}, we approximate $\mbf{H}^{\mrm{s}}$ for $\bsm{\chi} \in \mal{R}_q$ as
\begin{align}
	\mbf{H}^{\mrm{s}}(\bsm{\chi})  = \sum_{l=1}^{L^{\mrm{s}}} \beta_l^{\mrm{s}}(t)  \mbf a_N(\tau_l^{\mrm{s}}) \mbf b(\theta_l^{\mrm{s}},\phi_l^{\mrm{s}})^T + \bsm{\Delta}_{\bsm{\chi}},  \label{apr_H_s}
\end{align}
where $\beta_l^{\mrm{s}}(t)$ is the $l$-th effective complex coefficient with the path power being  $\rho_l^{\mrm{s}} \triangleq \mbb{E}[|\beta_l^{\mrm{s}}(t)|^2] $, and $\bsm{\Delta}_{\bsm{\chi}}$ is the representation error. 
Here, we express $\beta_l^{\mrm{s}}$ as a function of $t$, since the grid points rather than user poses are considered in stage I. The value of $L^{\mrm{s}}$ controls the trade-off between representation accuracy and construction complexity. $L^{\mrm{s}}$ can be less than $L$ since an effective path can approximate multiple physical paths with close delays and angles. The objective is to minimize the representation error over all poses, i.e.,
\begin{align}
	\min_{\bsm{\Phi}} \sum_{\bsm{\chi} \in \mal{R}_q} \left[ \left\|\sum_{l=1}^{L^{\mrm{s}}} \beta_l^{\mrm{s}}(t) \mbf a_N(\tau_l^{\mrm{s}}) \mbf b(\theta_l^{\mrm{s}},\phi_l^{\mrm{s}})^H - \mbf H^{\mrm{s}}(\bsm{\chi}) \right\|^2_F \right],  \label{eq:representation_error}
\end{align}
where $\bsm{\Phi} = \{ \tau_l^{\mrm{s}},\theta_l^{\mrm{s}},\phi_l^{\mrm{s}},\beta_l^{\mrm{s}}(t) \}$.

Recall that the quasi-static parameters $\{\tau_l^{\mrm{s}},\theta_l^{\mrm{s}},\phi_l^{\mrm{s}}\}_{l=1}^{L^{\mrm{s}}}$ are intended to capture the average channel characteristics within the $q$-th grid. Their estimation relies on a substantial volume of historical measurement data collected at the BS. Let $T$ be the number of historical measurements within grid $q$. Without loss of generality, the pilot symbol on each subcarrier is set to one. Assume that the cyclic prefix (CP) is longer than the maximum delay spread. After removing the CP and performing a discrete Fourier transform (DFT), the received signal is expressed as
\begin{align}
	\mbf Y^{\mrm{s}}(t)= \mbf{H}^{\mrm{BB},\mrm{s}}(t) + \mbf{H}^{\mrm{BB},\mrm{d}}(t) +  \mbf W_t, ~ t = 1,...,T, \label{eq:rx_signal_model}
\end{align}
where $\mbf W_t \in \mathbb{C}^{N \times M}$ is the AWGN matrix with i.i.d. entries $\sim \mathcal{CN}(0, \sigma_w^2)$. Substituting \eqref{eq:H_BB} and \eqref{apr_H_s} into \eqref{eq:rx_signal_model}, we obtain the received signal model for the $q$-th grid:
\begin{align}
	\mbf Y^{\mrm{s}}(t) = \sum_{l=1}^{L^{\mrm{s}}} \beta_l^{\mrm{s}}(t)  \mbf a_N(\tau_l^{\mrm{s}} + \epsilon(t)) \mbf b(\theta_l^{\mrm{s}},\phi_l^{\mrm{s}})^T + \widetilde{\mbf W}_t, \label{eq:grid_rx_model_matrix}
\end{align}
where the effective noise $\widetilde{\mbf W}_t \in \mathbb{C}^{N \times M}$ combines the AWGN, the representation error, and the dynamic channel component:
\begin{align}
	\widetilde{\mbf W}_t \triangleq  \bsm{\Delta}_{\bsm{\chi}} + \mbf{H}^{\mrm{BB, d}}(t)  + \mbf W_t.
\end{align}
The estimation of quasi-static parameters from historical data $\{\mbf Y^{\mrm{s}}(t)\}_{t=1}^T$ faces two challenges. First, since $\mbf Y^{\mrm{s}}(t),\forall t$ are collected from various user poses, the path coefficients $\beta_l^{\mrm{s}}(t)$ exhibit significant power spectral variation. This variation complicates the extraction of the common parameters $\{\tau_l^{\mrm{s}},\theta_l^{\mrm{s}},\phi_l^{\mrm{s}}\}$ from $\{\mbf Y^{\mrm{s}}(t)\}_{t=1}^T$. Second, each observation is affected by a different, unknown $\epsilon(t)$, which should be calibrated under non-informative priors of $\tau_l^{\mrm{s}}$ and $\epsilon(t)$. How to address these issues is detailed in Sec. \ref{sec:CKMM} and Sec. V.

\vspace{-0.2cm}
\subsection{Stage II: Estimation of Dynamic Parameters}

The dynamic parameter estimation aims to provide a user-specific channel parameters, which is crucial for PHY tasks such as channel estimation, beamforming, and data detection. Its construction must be performed with low latency and limited pilot overhead, leveraging real-time measurements such as DM-RS or SRS \cite{TS38211}. Hence, we employ a single OFDM symbol in each estimation period. For notation simplicity, we focus on a specific OFDM symbol and omit the time index. The baseband channel at a specific instant is expressed as
\begin{align}
	\mbf{H}^{\mrm{BB}} = & \sum_{l=1}^{L^{\mrm{s}}} \beta_l^{\mrm{s}} \mbf a_N(\tau_l^{\mrm{s}} + \epsilon) \mbf b(\theta_l^{\mrm{s}}, \phi_l^{\mrm{s}})^T \notag \\
	& + \sum_{l=1}^{L^{\mrm{d}}} \beta_{l}^{\mrm{d}} \mbf a_N(\tau_{l}^{\mrm{d}}) \mbf b(\theta_{l}^{\mrm{d}}, \phi_{l}^{\mrm{d}})^T + \bsm{\Delta}^{\mrm{BB}} , \label{eq:H_dyn_effective}
\end{align}
where the first term represents the channel component associated with the quasi-static scatterers, where $\{\tau_l^{\mrm{s}}, \theta_l^{\mrm{s}}, \phi_l^{\mrm{s}}\}$ are the same as those at stage I, and $\beta_l^{\mrm{s}}$ exhibits variations due to the change of pose. The second term represents the dynamic components, where the effective delay $\tau_l^{\mrm{d}}$ inherently includes the synchronization error $\epsilon$ which cannot be resolved from the measurement of a single OFDM symbol. The third term $\bsm{\Delta}^{\mrm{BB}}$ is the representation error.

To further reduce the overhead, consider that the pilot symbols are transmitted on a subset of the available subcarriers. Let $\mbf S \in \{0, 1\}^{P \times N}$ be a selection matrix that picks the $P$ pilot subcarriers ($P \le N$). Without loss of generality, the pilot symbols are set to ones. With $\mbf S$ and \eqref{eq:H_dyn_effective}, the received signal is
\begin{align}
	\mbf Y^{\mrm{d}} = & \mbf S \mbf{H}^{\mrm{BB}} + \widetilde{\mbf W} \notag \\
	 = &\sum_{l=1}^{L^{\mrm s}} \beta_l^{\mrm{s}} \tilde{\mbf a}_N(\tau_l^{\mrm{s}} + \epsilon) \mbf b(\theta_l^{\mrm{s}}, \phi_l^{\mrm{s}})^T \notag \\
	& + \sum_{l=1}^{L^{\mrm d}} \beta_{l}^{\mrm{d}} \tilde{\mbf a}_N(\tau_{l}^{\mrm{d}}) \mbf b(\theta_{l}^{\mrm{d}}, \phi_{l}^{\mrm{d}})^T  + \widetilde{\mbf W}, \label{eq:dyn_rx_model}
\end{align}
where $\tilde{\mbf a}_N(\cdot) \triangleq \mbf S \mbf a_N(\cdot)$ is the compressed steering vector, and $\widetilde{\mbf W}$ is the effective noise matrix. From the noisy observation $\mbf Y^{\mrm{d}}$, we aim to estimate $\{\{\beta_l^{\mrm{s}}\}, \epsilon, \{\beta_l^{\mrm{d}}, \tau_l^{\mrm{d}}, \theta_l^{\mrm{d}}, \phi_l^{\mrm{d}}\}\}$. The main chanllenge is that the dynamic parameters vary rapidly, while the estimator should remain low-complexity and accurate under limited pilot overhead. To this end, we exploit the quasi-static parameters obtained in stage~I as strong informative priors, which substantially shrinks the search space.  The detailed solution is provided in Sec. IV and Sec. VI.









\section{Bayesian Inference Framework for Dynamic CKM Construction} \label{sec:CKMM}

In this section, we formulate the dynamic CKM construction problem into a sequential Bayesian inference framework. 


\subsection{Probability Model for Quasi-Static Parameter Estimation}

In the quasi-static parameter estimation, we infer the set of grid-level quasi-static channel parameters $ \{\tau_l^{\mrm{s}}, \theta_l^{\mrm{s}}, \phi_l^{\mrm{s}}\}_{l=1}^{L^{\mrm{s}}}$ and calibrate the synchronization errors $\{\epsilon(t)\}_{t=1}^T$. We adopt a Bayesian approach, beginning with the specification of prior distributions. For the circular variables $\vartheta_l^{\mrm{s}} \in \bsm{\varTheta}^{\mrm{s}} \triangleq \{\tau_l^{\mrm{s}}, \theta_l^{\mrm{s}}, \phi_l^{\mrm{s}}\}_{l=1}^{L^{\mrm{s}}}$, we employ the von Mises distribution \cite{mardia2009directional}  as 
\begin{align}
	p(\vartheta_l^{\mrm{s}}) = \mal{VM}(\vartheta_l^{\mrm{s}}; \mu_l^{\mrm{s}}, \kappa_l^{\mrm{s}}),  \label{delay_angle_VM_s}
\end{align}
where $\mu_l^{\mrm{s}}$ and $\kappa_l^{\mrm{s}}$ are the mean and concentration parameters, respectively. When no prior knowledge is available, we set $\kappa_l^{\mrm{s}} = 0$, which reduces the von Mises distribution to a uniform distribution over $[0, 2\pi)$. Similarly, $\epsilon(t)$ is modeled as $p(\epsilon(t)) = \mal{VM}(\epsilon(t); \mu_\epsilon, \kappa_\epsilon)$.

For the complex path coefficient $\beta_l^{\mrm{s}}(t)$, we adopt a complex Gaussian prior as
\begin{align}
	p(\beta_l^{\mrm{s}}(t)) = \mal{CN}(\beta_l^{\mrm{s}}(t);0,\rho_l^{\mrm{s}}), \label{path_coeff_s}
\end{align}
where $\rho_l^{\mrm{s}}$ is the path power to be estimated.

We define the set of all parameters to be inferred as $\bsm{\xi}^{\mrm{s}} \triangleq \{ \{\tau_l^{\mrm{s}}, \theta_l^{\mrm{s}}, \phi_l^{\mrm{s}}\}_{l=1}^{L^{\mrm{s}}}, \{\epsilon(t)\}_{t=1}^T, \{\bsm{\beta}^{\mrm{s}}(t)\}_{t=1}^T \}$, where $\bsm{\beta}^{\mrm{s}}(t) = [\beta_1^{\mrm{s}}(t), \dots, \beta_{L^{\mrm{s}}}^{\mrm{s}}(t)]^T$. Let $\mbf Y^{\mrm{s}} \triangleq [\mbf Y^{\mrm{s}}(1), \dots, \mbf Y^{\mrm{s}}(T)]$ be the collection of received signals over $T$ time slots. Given $\mbf Y^{\mrm{s}}$, the posterior distribution of $\bsm{\xi}^{\mrm{s}}$ is given by
\begin{align}
	p(\bsm{\xi}^{\mrm{s}} | \mbf Y^{\mrm{s}}) \propto &  \prod_{t=1}^T p(\mbf Y^{\mrm{s}}(t)| \bsm{\xi}^{\mrm{s}})  \prod_{l=1}^{L^{\mrm{s}}} p(\tau_l^{\mrm{s}}) p(\theta_l^{\mrm{s}}) p(\phi_l^{\mrm{s}})  \notag \\
	& \times  \prod_{t=1}^T p(\epsilon(t)) \prod_{l=1}^{L^{\mrm{s}}} p(\beta_l^{\mrm{s}}(t)) . \label{eq:posterior_static}
\end{align}

\subsection{Probability Model for Dynamic Parameter Estimation}

In the dynamic parameter estimation, we aim to refine the quasi-static components and simultaneously estimate the dynamic components. For the quasi-static channel parameters $\vartheta_l^{\mrm{s}} \in \{\tau_l^{\mrm{s}}, \theta_l^{\mrm{s}}, \phi_l^{\mrm{s}}\}$, their priors are obtained from the quasi-static parameter estimation stage:
\begin{align}
	p(\vartheta_l^{\mrm{s}}) \propto \int p(\bsm{\xi}^{\mrm{s}} | \mbf Y^{\mrm{s}}) d\bsm{\xi}^{\mrm{s}} \setminus \vartheta_l^{\mrm{s}}, \label{static_prior}
\end{align}
where $p(\bsm{\xi}^{\mrm{s}} | \mbf Y^{\mrm{s}})$ is from \eqref{eq:posterior_static}. For the complex coefficient $\beta_l^{\mrm{s}}$, we use the estimated power $\hat{\rho}_l^{\mrm{s}}$ (from stage I) as $p(\beta_l^{\mrm{s}} | \hat{\rho}_l^{\mrm{s}}) = \mal{CN}(\beta_l^{\mrm{s}};0,\hat{\rho}_l^{\mrm{s}})$.

For the dynamic parameters $\vartheta_l^{\mrm{d}} \in \bsm{\varTheta}^{\mrm{d}} \triangleq \{\tau_l^{\mrm{d}}, \theta_l^{\mrm{d}}, \phi_l^{\mrm{d}}\}_{l=1}^{L^{\mrm{d}}}$, we adopt non-informative von Mises priors similar to \eqref{delay_angle_VM_s}.
The key distinction lies in modeling the complex coefficients $\{\beta_l^{\mrm{d}}\}$. We use a Bernoulli-Gaussian (BG) prior:
\begin{align}
	p(\beta_l^{\mrm{d}}) = (1 - \lambda_l^{\mrm{d,pri}})\, \delta(\beta_l^{\mrm{d}}) + \lambda_l^{\mrm{d,pri}}\, \mal{CN}(\beta_l^{\mrm{d}};0,v_l^{\mrm{d,pri}}),
\end{align}
where $\lambda_l^{\mrm{d,pri}}$ is the prior probability of the path being active, and $v_l^{\mrm{d,pri}}$ is its prior power when active.

Let $\bsm{\xi}$ be the collection of all parameters to be estimated. Given the received signal $\mbf Y^{\mrm{d}}$, the posterior distribution is
\begin{align}
	p(\bsm{\xi} | \mbf Y^{\mrm{d}}) \propto ~ & p(\mbf Y^{\mrm{d}}| \bsm{\xi}) \prod_{l=1}^{L^{\mrm{s}}} p(\beta_l^{\mrm{s}} | \hat{\rho}_l^{\mrm{s}}) p(\vartheta_l^{\mrm{s}}) \notag \\
	& \times p(\epsilon) \prod_{l=1}^{L^{\mrm{d}}} p(\beta_l^{\mrm{d}}) p(\tau_l^{\mrm{d}}) p(\theta_l^{\mrm{d}}) p(\phi_l^{\mrm{d}}). \label{eq:posterior_dynamic}
\end{align}

\subsection{Approximate Bayesian Inference}

\vspace{-0.2 cm}
\begin{figure}[htbp]
	\centering
	\includegraphics[width=0.5\textwidth]{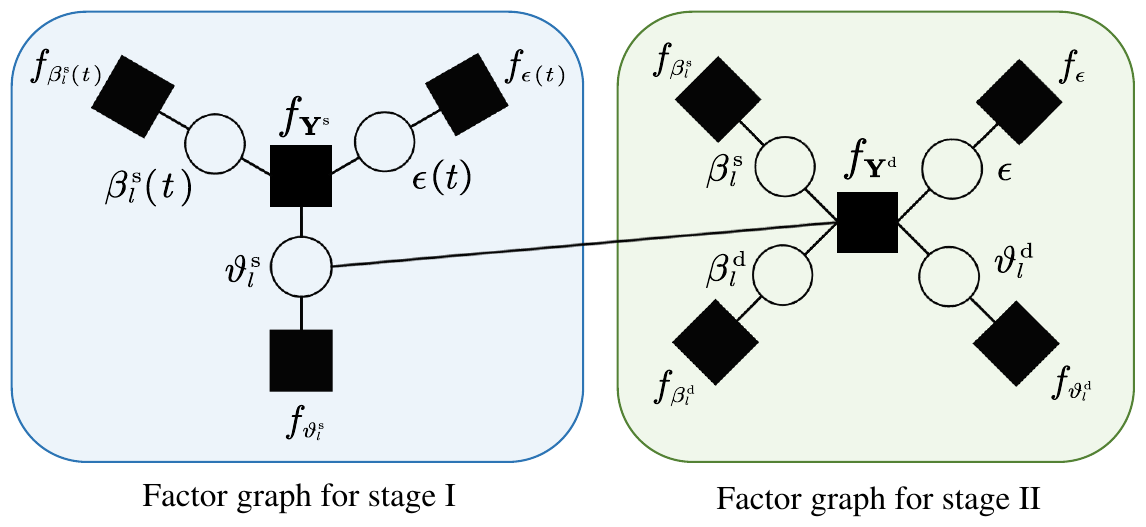}
	\caption{Factor graph representation for dynamic CKM construction. Left and right subgraphs correspond to probability factorization \eqref{eq:posterior_static} and \eqref{eq:posterior_dynamic}, respectively.}
	\label{fig:FG}
\end{figure}
\vspace{-0.2 cm}

The Bayesian optimal estimator requires computing the marginal posteriors from \eqref{eq:posterior_static} and \eqref{eq:posterior_dynamic}, which is computationally intractable due to the high-dimensional integrals involved. To address this, we employ approximate Bayesian inference \cite{BP,Unify_2,HVMP}. To begin with, we represent the posterior distributions (\eqref{eq:posterior_static} and \eqref{eq:posterior_dynamic}) with the factor graph as in Fig. \ref{fig:FG}. Circles represent variable nodes, corresponding to the parameters to be estimated;
Squares represent factor nodes, corresponding to the probability factors in \eqref{eq:posterior_static} and \eqref{eq:posterior_dynamic}. To simplify the notation, we denote the variable nodes and factor nodes by $v$ and $f$, respectively, and group all variables connected to a factor node $f$ into a vector $\bsm u_f$. At variable and factor nodes, we introduce \emph{beliefs} $b_{v}(v)$ and $b_f(\bsm u_f)$, respectively. The Bethe approximation of the posterior distribution is \cite{BP}
\begin{align}
	b(\bsm{\xi}) = {\prod_{f \in \mal{F}} b_f(\bsm u_f) } / \left({\prod_{v\in\mal{V}} [ b_{v}(v)]^{A_v-1}} \right), \label{b_fac}
\end{align}
where $\mal{V}$ is the set of variable nodes, $\mal{F}$ is the set of factor nodes, and $A_v$ is the number of variables connected to node $v$. We aim to optimize $b(\bsm{\xi})$ to minimize the Kullback-Leibler (KL) divergence:
\begin{align}
	\mrm{KL} \big[ b(\bsm{\xi}) \| p(\bsm{\xi} | \mbf Y) \big] = \mathbb{E}_{b(\bsm{\xi})} \left[ \log \frac{b(\bsm{\xi})}{p(\bsm{\xi} | \mbf Y)} \right]. \label{KL_def}
\end{align}
In stage I and II, we have $\mbf Y = \mbf Y^{\mrm{s}}$ and $\mbf Y = \mbf Y^{\mrm{d}}$, respectively.

We then introduce matching constraints between $b_{v}(v)$ and $b_f(\bsm u_f)$ to ensure consistency:
	\begin{equation}
		\label{b_v_cons1}
		\begin{aligned}
		\mathbb{E}_{b_v(v)}[\cos(v)] &= \mathbb{E}_{\int b_f(\bsm u_f) d\bsm u_f \setminus v}[\cos(v)], \quad \text{if } v \text{ is circular}, \\
		\mathbb{E}_{b_v(v)}[\sin(v)] &= \mathbb{E}_{\int b_f(\bsm u_f) d\bsm u_f \setminus v}[\sin(v)], \quad \text{if } v \text{ is circular}, \\
		\mathbb{E}_{b_v(v)}[v] &= \mathbb{E}_{\int b_f(\bsm u_f) d\bsm u_f \setminus v}[v], \quad \text{if } v \text{ is Gaussian}, \\
		\mathbb{E}_{b_v(v)}[|v|^2] &= \mathbb{E}_{\int b_f(\bsm u_f) d\bsm u_f \setminus v}[|v|^2], \quad \text{if } v \text{ is Gaussian}, \\
		\mathbb{P}_{b_v(v)}(v=0) &= \mathbb{P}_{\int b_f(\bsm u_f) d\bsm u_f \setminus v}(v=0), \quad \text{if } v \text{ is BG},
		\end{aligned}
	\end{equation}
plus the first and the second moment matchings for the BG variable.

For a tractable inference process, we introduce belief decompositions. In stage I, the belief at the likelihood factor node $f_{\mbf Y^{\mrm{s}}}$ is decomposed as
\begin{align}
	b_{f_{\mbf Y^{\mrm{s}}}}(\bsm u_f) = \prod_{t=1}^T b_{f_{\mbf Y^{\mrm{s}}}}(\bsm{\beta}^{\mrm{s}}(t),\epsilon(t) ) \prod_{v \in \bsm{\varTheta}^{\mrm{s}} } b_{f_{\mbf Y^{\mrm{s}}}}(v). \label{b_f_cons1} 
\end{align} 
In stage II, the belief $b_{f_{\mbf Y^{\mrm{d}}}}(\bsm u_f)$ is decomposed as
\begin{align}
	b_{f_{\mbf Y^{\mrm{d}}}}(\bsm u_f) = b_{f_{\mbf Y^{\mrm{d}}}}(\bsm{\beta}^{\mrm{s}},\epsilon) \prod_{l=1}^{L^{\mrm{d}}} b_{f_{\mbf Y^{\mrm{d}}}}(\beta_l^{\mrm{d}}) \prod_{v \in \bsm{\varTheta}^{\mrm{s}} \cup \bsm{\varTheta}^{\mrm{d}}} b_{f_{\mbf Y^{\mrm{d}}}}(v), \label{b_f_cons2} 
\end{align} 

Given \eqref{b_v_cons1}-\eqref{b_f_cons2}, the minimization of the Bethe free energy \eqref{KL_def} yields a set of fixed-point equations as
	\begin{subequations}
		\label{HMP_b}
		\begin{align}
			& b_{f_v}(v) \propto  {f}(v) \prod_{v \in \mal{N}(f_v) } \mal{M}_{v \to f_v}(v), \label{HMP_bf} \\
			& b_{f_{\mbf Y}}(\bsm u_{f,i}) \propto \exp \Big( \mathbb{E}_{b(\bsm u_f \backslash \bsm u_{f,i})} \left[ \ln {f_{\mbf Y}}(\bsm u_f) \right] \Big) \prod_{v \in \bsm u_{f,i} } \mal{M}_{v \to f_{\mbf Y}}(v), \label{HMP_bfi} \\
			& b_{v}(v) \propto \prod_{f \in \mal{N}(v)} \mal{M}_{f \to v}(v), \label{HMP_bv}
		\end{align}
	\end{subequations}
	where $\mal{N}(f)$ is the set of neighboring variable nodes of factor $f$, $\mal{N}(v)$ is the set of neighboring factor nodes of variable $v$, and
	\begin{subequations}
		\label{HMP_mp}
		\begin{align}
			& \mal{M}_{v \to f}(v) \propto \prod_{\tilde{f} \in \mal{N}(v) \backslash f } \mal{M}_{ \tilde{f} \to v} (v), \label{mp_a} \\
			& \mal{M}_{f_v \to v} (v) \propto \mal{P} \left( b_{f_v}(v)  \right)/ \mal{M}_{v \to f}(v), \label{mp_b} \\
			& \mal{M}_{f_{\mbf Y} \to v} (v) \propto \mal{P} \left(\int b_{f_{\mbf Y}}(\bsm u_{f_{\mbf Y},i}) d \bsm u_{f_{\mbf Y},i} \backslash v  \right)/ \mal{M}_{v \to f_{\mbf Y}}(v).  \label{mp_c} 
		\end{align}
	\end{subequations}
	Here, $\mal{P}(\cdot)$ is a projection operator that projects a belief onto the specified family of distributions (von Mises, Gaussian, or BG) by matching the constraints in \eqref{b_v_cons1}.

	The proof of \eqref{HMP_b}-\eqref{HMP_mp} mainly follows \cite[Appendix A]{Wenjun_HVMP}, and is omitted for brevity. A key difference lies in the constraints \eqref{b_v_cons1} for circular, Gaussian, and BG variables, which make the projected messages in \eqref{mp_b}-\eqref{mp_c} remain consistent with their respective prior distribution types. Furthermore, we introduce tailored belief decompositions for the dynamic CKM construction. Specifically, in \eqref{b_f_cons1}, the joint belief $b_{f_{\mbf Y^{\mrm{s}}}}(\bsm{\beta}^{\mrm{s}}(t),\epsilon(t))$ is introduced to enable joint estimation of $\bsm{\beta}^{\mrm{s}}(t)$ and $\epsilon(t)$, which facilitates high-performance group delay calibration even with non-informative priors. In \eqref{b_f_cons2}, the marginal beliefs $b_{f_{\mbf Y^{\mrm{d}}}}(\beta_l^{\mrm{d}}), \forall l$ are introduced to enable path-wise low-complexity estimation of $\beta_l^{\mrm{d}},\forall l$.

Next, we show how the introduced constraints \eqref{b_v_cons1} facilitate the belief and message updates \eqref{HMP_b}-\eqref{HMP_mp}. Specifically, each variable node in Fig. \ref{fig:FG} is connected to its prior factor node $f_v$ and the likelihood factor node $f_{\mbf Y}$. Thus, \eqref{mp_a} simplifies to
\begin{subequations}
	\label{mp_a_simplify}
	\begin{align}
	& \mal{M}_{v \to f_{\mbf Y}}(v) = \mal{M}_{ f_v \to v} (v), \label{mp_a_simplify_1} \\
	& \mal{M}_{v \to f_v}(v) = \mal{M}_{ f_{\mbf Y} \to v} (v). \label{mp_a_simplify_2}
\end{align}
\end{subequations}
A typical iteration starts by initializing $\mal{M}_{ f_v \to v} (v) = p(v)$. Then, we apply \eqref{mp_b} to obtain
\begin{align}
	\mal{M}_{f_{\mbf Y} \to v} (v) = \mal{P} \left(\int b_{f}(\bsm u_{f,i}) d \bsm u_{f,i} \backslash v \right) / p(v). \label{mp_b_simplify}
\end{align}
Recall that the projection operator $\mal{P}(\cdot)$ is consistent with the prior $p(v)$. Besides, Gaussian, Von Mises, or BG distribution is closed under multiplication/division operations, or more specifically, the division of two distributions of the same type results in another distribution of the same type. Therefore, $\mal{M}_{f_{\mbf Y} \to v} (v)$ in \eqref{mp_b_simplify} follows the same type of distribution as the prior $p(v)$. Then, by substituting \eqref{HMP_bf} into \eqref{mp_b}, we obtain
\begin{subequations}
	\begin{align}
	\mal{M}_{f_{\bsm{v}} \to v}(v) 
	& = {\mal{P}( p(v) \mal{M}_{f_{\mbf Y} \to v} (v) )} / {\mal{M}_{f_{\mbf Y} \to v} (v) }, \label{Sec4_Mf_v} \\
	& = p(v). \label{Sec4_Mf_tau}
\end{align}
\end{subequations}
We then substitute \eqref{mp_a_simplify_1} and \eqref{Sec4_Mf_tau} into \eqref{HMP_bfi}, and substitute \eqref{mp_a_simplify_1} and \eqref{mp_c} into \eqref{HMP_bv}, which yields
\begin{subequations}
	\label{HMP_b_simplify}
	\begin{align}
	& b_f(\bsm u_{f,i}) \propto \exp \Big( \mathbb{E}_{b_f(\bsm u_f \backslash \bsm u_{f,i})} \left[ \ln {f}(\bsm u_f) \right] \Big) \prod_{v \in \bsm u_{f,i} } p(v) \label{b_fi_final},  \\
	& b_{v}(v) \propto \mal{P} \left( \int b_{f}(\bsm u_{f,i}) d \bsm u_{f,i} \backslash v   \right). \label{b_v_final}
\end{align}
\end{subequations}
Here, we show that with the constraint \eqref{b_v_cons1}, \eqref{HMP_b}-\eqref{HMP_mp} equivalate to \eqref{HMP_b_simplify}. By computing the expectation of $v$ w.r.t. $b_v(v)$, we obtain the estimate of channel parameters.

\section{Algorithm Design for Quasi-Static Parameter Estimation}


\subsection{Beliefs of Path Delays and Angles}
Let's consider the belief for the path delay $\tau_l^{\mrm s}$. We apply \eqref{b_fi_final} to obtain 
\begin{align}
	b_{f_{\mbf Y}}(\tau_l^{\mrm s}) = g(\tau_l^{\mrm s}) p(\tau_l^{\mrm s}), \label{eq:Sec4_bf_Yt_tau}
\end{align}
where 
\begin{align}
	& \ln g(\tau_l^{\mrm s}) \notag \\
	& \propto \mathbb{E}_{\bsm{\xi} \setminus \tau_l^{\mrm s}} [ -\frac{1}{\sigma_w^2} \sum_t \|\mbf Y_t - \sum_{l} \beta_l^{\mrm{s}}(t) \mbf a_N(\tau_l^{\mrm s} + \epsilon(t)) \mbf b(\theta_l^{\mrm s}, \phi_l^{\mrm s})^T \|_F^2 ]. \notag
\end{align}
To derive $g(\tau_l^{\mrm s})$, we introduce the residual signal for the $l$-th path at time $t$ as
\begin{equation}
	\hat{\mathbf{R}}_{t,l} \triangleq \mathbf{Y}^{\mrm s}(t) - \sum_{j \neq l} \hat{\beta}_j^{\mrm{s}}(t) \hat{\mbf a}_j \hat{\mathbf{b}}_j^T, \label{residual}
\end{equation}
where $\hat{\beta}_j^{\mrm{s}}(t) = \mathbb{E}[\beta_j^{\mrm{s}}(t)]$, $\hat{\mbf a}_j = \mathbb{E}[\mathbf{a}_N({\tau}_j^{\mrm s} + {\epsilon}(t))]$, and $\hat{\mathbf{b}}_j = \mathbb{E}[\mathbf{b}({\theta}_j^{\mrm s}, {\phi}_j^{\mrm s})]$. Given \eqref{residual}, the term in $\ln g(\tau_l^{\mrm s})$ that depends on $\tau_l^{\mrm s}$ is expressed as
\begin{align}
	&  \operatorname{Re}  \left\{  \frac{2}{\sigma_w^2} \mbb E \left[  \mathrm{Tr} \left[ \sum_t \left( \beta_l^{\mrm{s}}(t) \mathbf{a}_N(\tau_l^{\mrm{s}} + \epsilon(t))\, \mathbf{b}_l^T \right)^H \hat{\mathbf{R}}_{t,l} \right] \right]  \right\} \notag \\
	& = \operatorname{Re}  \left\{ \frac{2}{\sigma_w^2} \sum_t \hat{\beta}_l^{\mrm{s}}(t)^*\, \mathbf{a}_N(\tau_l^{\mrm{s}})^H\, \mrm{diag}( \mbb{E}[\mbf a_N(\epsilon(t)) ] )^H\, \hat{\mathbf{R}}_{t,l}\, \hat{\mathbf{b}}_l^* \right\} , \notag
\end{align}
We then write $g(\tau_l^{\mrm{s}})$ as
\begin{align}
	g(\tau_l^{\mrm{s}}) &\propto \exp \left( \operatorname{Re} \left\{ \boldsymbol{\eta}_{g(\tau_l^{\mrm{s}})}^H\, \mathbf{a}_N(\tau_l^{\mrm{s}}) \right\} \right), \label{eq:Sec4_g_tau} \\
	\boldsymbol{\eta}_{g(\tau_l^{\mrm{s}})} & \triangleq \frac{2}{\sigma_w^2}\,  \sum_t \hat{\beta}_l^{\mrm{s}}(t)^*\,  \mrm{diag}( \mbb{E}[\mbf a_N(\epsilon(t)) ] )\, \hat{\mathbf{R}}_{t,l}\, \hat{\mathbf{b}}_l^*. \label{eq:Sec4_eta_g_tau}
\end{align}
We rewrite $p(\tau_l^{\mrm s}) \propto \exp\left( \operatorname{Re} \left\{ \bsm{\eta}_{p(\tau_l^{\mrm s})}^H \mbf{a}_N(\tau_l^{\mrm s}) \right\} \right)$, where $\bsm{\eta}_{p(\tau_l^{\mrm s})} = [0,\kappa_l^{\mrm{s}} e^{j \mu_l^{\mrm{s}}},0,...,0]$. By substituting $g(\tau_l^{\mrm s})$ and $p(\tau_l^{\mrm s})$ into \eqref{eq:Sec4_bf_Yt_tau}, we obtain
\begin{align}
	b_{f_{\mbf Y}}(\tau_l^{\mrm s}) \propto \exp \left( \operatorname{Re} \left\{ (\bsm{\eta}_{g(\tau_l^{\mrm s})} + \bsm{\eta}_{p(\tau_l^{\mrm s})})^H \mbf a_N(\tau_l^{\mrm s}) \right\} \right). \label{eq:bf_tau_final}
\end{align}
To estimate $\tau_l^{\mrm s}$, we can perform a grid search over the domain $[0, 2\pi)$, i.e.,
\begin{align}
	\hat{\tau}_{l,\text{coarse}}^{\mrm s} = \argmax_{\omega \in \mathcal{G}} \operatorname{Re} \left\{ \bsm{\eta}_{\tau_l^{\mrm s}}^H \mbf a_N(\omega) \right\},
\end{align}
where $\mathcal{G}$ is a fine-grained grid of candidate values, and $\bsm{\eta}_{\tau_l^{\mrm s}} = \bsm{\eta}_{g(\tau_l^{\mrm s})} + \bsm{\eta}_{p(\tau_l^{\mrm s})}$. This grid search can be efficiently implemented using an $N$-point FFT with grid offsetting. Specifically, we define the number of offsetting as $Q$ and perform a search as
\begin{align} 
	(n^*,q^*) = \argmax_{n,q} ~ \mbf a_N^H\left(2\pi\frac{n}{N}\right) \left( \bsm{a}_N\left(2\pi\frac{q}{QN} \right) \odot \bsm{\eta}_{\tau_l^{\mrm s}} \right), \label{eq:FFT_search}
\end{align}
where $\mbf a_N\left(2\pi\frac{n}{N}\right)$ is the $n$-th column of the $N \times N$ DFT matrix, and $q\in\{1,..,Q\}$. The pair $(n^*,q^*)$ corresponds to the estimate $\hat{\tau}_{l,\text{coarse}}^{\mrm s} = 2\pi \left(  \frac{n^*-1}{N} +  \frac{q^*-1}{NQ} \right)$. $\hat{\tau}_{l,\text{coarse}}^{\mrm s}$ is then refined using Newton's method on $f(\omega) = \operatorname{Re} \{ \bsm{\eta}_{\tau_l^{\mrm s}}^H \mbf a_N(\omega) \}$, iterating $\omega_{k+1} = \omega_k - f'(\omega_k)/f''(\omega_k)$ to find the refined $\hat{\tau}_{l}^{\mrm s}$. Note that a Newton update of $\tau_l^{\mrm s}$ is equivalent to a second-order Taylor expansion of $\ln b_{f_{\mbf Y_t}}(\tau_l^{\mrm s})$ at $\tau_l^{\mrm s} = \hat{\tau}_{l}^{\mrm s}$. This expansion results in a Gaussian approximation of $b_{f_{\mbf Y_t}}(\tau_l^{\mrm s})$ with mean $\hat{\tau}_{l}^{\mrm s}$ and variance $(\sigma_{\tau_l^{\mrm s}})^2 = -1/\left( d^2 \ln b_{f_{\mbf Y_t}}(\tau_l^{\mrm s})/d(\tau_l^{\mrm s})^2 |_{\tau_l^{\mrm s}=\hat{\tau}_{l}^{\mrm s}} \right)$. By using the similarity between Gaussian and von Mises distributions \cite[Sec. 3.5]{mardia2009directional}, we obtain
\begin{align}
	b_{f_{\mbf Y_t}}(\tau_l^{\mrm s}) \approx \mal{VM}(\tau_l^{\mrm s}; \hat{\tau}_{l}^{\mrm s}, \kappa_{\tau_l^{\mrm s}}), \label{eq:bf_tau_approx}
\end{align}
with $\kappa_{\tau_l^{\mrm s}} = \mal{A}^{-1}(\exp(0.5/(\sigma_{\tau_l^{\mrm s}})^2))$, where $\mal{A}^{-1}(\cdot)$ is the inverse of $\mal{A}(x) = I_1(x)/I_0(x)$.

The belief update of $\theta_l^{\mrm s}$ is similar to that of $\tau_l^{\mrm s}$. We obtain $b_{f_{\mbf Y_t}}(\theta_l^{\mrm s}) = g(\theta_l^{\mrm s}) p(\theta_l^{\mrm s})$
with
\begin{align}
	g(\theta_l^{\mrm s})
	&\propto \exp \left( \operatorname{Re} \left\{ \boldsymbol{\eta}_{g(\theta_l^{\mrm s})}^H \mbf a_{M_1}(\theta_l^{\mrm s}) \otimes \hat{\mbf a}_{M_2} \right\} \right), \label{g_theta_s} \\
	\boldsymbol{\eta}_{g(\theta_l^{\mrm s})} &\triangleq \frac{2}{\sigma_w^2}\, \sum_t \hat{\beta}_l^{\mrm{s}}(t)^*  \hat{\mathbf{R}}_{t,l}^T\, \hat{\mathbf{a}}_l^*, \label{eta_theta_s} 
\end{align}
where $\hat{\mbf a}_{M_2} = \mbb{E}[ {\mbf a}_{M_2}(\phi_l^{\mrm s})]$. Then, we perform FFT-based search and Newton's method to obtain the estimate $\hat{\theta}_l^{\mrm s}$. The belief $b_{f_{\mbf Y_t}}(\theta_l^{\mrm s})$ is approximated as $\mal{VM}(\theta_l^{\mrm s}; \hat{\theta}_l^{\mrm s}, \kappa_{\theta_l^{\mrm s}})$, where $\kappa_{\theta_l^{\mrm s}}$ is computed similarly as $\kappa_{\tau_l^{\mrm s}}$. For $\phi_l^{\mrm s}$, it is symmetric to that of $\theta_l^{\mrm s}$ in the signal model \eqref{eq:grid_rx_model_matrix}. We follow the same process to obtain $b_{f_{\mbf Y_t}}(\phi_l^{\mrm s}) \approx \mal{VM}(\phi_l^{\mrm s}; \hat{\phi}_l^{\mrm s}, \kappa_{\phi_l^{\mrm s}})$.

\subsection{Beliefs of Synchronization Errors and Path Coefficients}

To facilitate belief computation, we vectorize the received signal model in \eqref{eq:grid_rx_model_matrix}. Let $\mbf y(t) \triangleq \mrm{vec}(\mbf Y^{\mrm{s}}(t))$, $\bsm{\beta}^{\mrm s}(t) \triangleq [\beta_1^{\mrm s}(t), \dots, \beta_{L^{\mrm s}}^{\mrm s}(t)]^T$, and $\tilde{\mbf w}_t \triangleq \mrm{vec}(\widetilde{\mbf W}_t)$. The model becomes
\begin{align}
	\mbf y(t) = \mbf A(\bsm{\vartheta}^{\mrm s}, \epsilon(t)) \bsm{\beta}^{\mrm s}(t) + \tilde{\mbf w}_t,
\end{align}
where $\bsm{\vartheta}^{\mrm s} \triangleq \{\tau_l^{\mrm s}, \theta_l^{\mrm s}, \phi_l^{\mrm s}\}_{l=1}^{L^{\mrm s}}$ and $\mbf A(\bsm{\vartheta}^{\mrm s}, \epsilon(t))$ is the dictionary matrix given by
\begin{align}
	\mbf A(\bsm{\vartheta}^{\mrm s}, \epsilon(t)) = [\dots,\mbf b(\theta_l^{\mrm s},\phi_l^{\mrm s}) \otimes \mbf a_N(\tau_l^{\mrm s} + \epsilon(t)), \dots].
\end{align}

Since the inference for $(\bsm{\beta}^{\mrm s}(t), \epsilon(t))$ is identical for each time slot $t$, we drop the index $t$ for notational simplicity in the following derivation. The belief of $(\bsm{\beta}^{\mrm s}, \epsilon)$ at the factor node $f_{\mbf Y}$ is given by
\begin{align}
	\ln b_{f_{\mbf Y}}(\bsm{\beta}^{\mrm s} , \epsilon) & \propto \mathbb{E}_{\bsm{\xi} \setminus \{ \bsm{\beta}^{\mrm s}, \epsilon \}} \left[ \ln f_{\mbf Y}(\bsm{\xi}) \right] + \ln p(\bsm{\beta}^{\mrm s}) + \ln p(\epsilon)  \notag \\
	& \propto g(\bsm{\beta}^{\mrm s},\epsilon) + f(\epsilon) + \ln p(\epsilon), \label{b_fYt_beta_epsilon}
\end{align}
The terms $g(\cdot)$ and $f(\cdot)$ arise from expanding the log-likelihood term. After some algebraic manipulations, we have
\begin{subequations}
	\begin{align}
		g(\bsm{\beta}^{\mrm s},\epsilon) &= - (\bsm{\beta}^{\mrm s} - \bsm{\mu}(\epsilon))^H \bsm{\Sigma}(\epsilon)^{-1} (\bsm{\beta}^{\mrm s} - \bsm{\mu}(\epsilon)), \label{g_term} \\
		f(\epsilon) &= \sigma_w^{-2}\mbf y^H {\mbf A}(\epsilon) ({\mbf \Gamma} +\sigma_w^2 \bsm{\Lambda}^{-1}  )^{-1} {\mbf A}(\epsilon)^H \mbf y, \label{fD}
	\end{align}	
\end{subequations}
with ${\mbf A}(\epsilon) \! = \! \mathbb{E}_{\bsm{\vartheta}^{\mrm{s}}}[\mbf A(\bsm{\vartheta}^{\mrm s}, \epsilon)]$, ${\mbf \Gamma} = \mathbb{E}_{\bsm{\vartheta}^{\mrm{s}}}[\mbf A(\bsm{\vartheta}^{\mrm s}, \epsilon)^H \mbf A(\bsm{\vartheta}^{\mrm s}, \epsilon)]$, $\bsm{\Lambda}=\mrm{diag}([\hat{\rho}_1^{\mrm{s}},..., \hat{\rho}_{L^{\mrm{s}}}^{\mrm{s}}])$, $\bsm{\mu}(\epsilon) = (\sigma_w^{-2} {\mbf \Gamma} +  \bsm{\Lambda}^{-1})^{-1} {\mbf A}(\epsilon)^H \mbf y$, $\bsm{\Sigma}(\epsilon) = (\sigma_w^{-2}{\mbf \Gamma} + \bsm{\Lambda}^{-1})^{-1}$, and ${\mbf \Gamma} = \mathbb{E}_{\bsm{\vartheta}^{\mrm{s}}}[\mbf A(\bsm{\vartheta}^{\mrm s}, \epsilon)^H \mbf A(\bsm{\vartheta}^{\mrm s}, \epsilon)]$. Note that ${\mbf \Gamma}$ is irrelavant to $\epsilon$. To see this, we can write $\mbf A(\bsm{\vartheta}^{\mrm s}, \epsilon)  = \mbf D(\epsilon) \mbf A(\bsm{\vartheta}^{\mrm{s}},0)$, where $\mbf D(\epsilon) = \mrm{diag}(\mbf a_N(\epsilon)) \otimes \mbf I_M$. Then, we have ${\mbf \Gamma} = \mathbb{E}_{\bsm{\vartheta}^{\mrm{s}}}[ \mbf A(\bsm{\vartheta}^{\mrm{s}},0)^H  \mbf D(\epsilon)^H  \mbf D(\epsilon) \mbf A(\bsm{\vartheta}^{\mrm{s}},0) ]= \mathbb{E}_{\bsm{\vartheta}^{\mrm{s}}}[ \mbf A(\bsm{\vartheta}^{\mrm{s}},0)^H \mbf A(\bsm{\vartheta}^{\mrm{s}},0) ]$.

The term $g(\bsm{\beta}^{\mrm s},\epsilon)$ in \eqref{g_term} is quadratic in $\bsm{\beta}^{\mrm s}$ (given $\epsilon$), implying a Gaussian belief for $\bsm{\beta}^{\mrm s}$. By marginalizing out $\bsm{\beta}^{\mrm s}$ from \eqref{b_fYt_beta_epsilon}, we obtain $b_{f_{\mbf Y}}(\epsilon) \propto \exp(f(\epsilon)) p(\epsilon)$. We now show that $f(\epsilon)$ can be expressed into a form suitable for spectral estimation. Specifically, we express ${\mbf A}(\epsilon)^H \mbf y$ as
\begin{align}
	{\mbf A}(\epsilon)^H \mbf y & = \mbf A(0)^H \mbf D(\epsilon)^H \mbf y = \mbf B \mbf a_N(\epsilon)^*, \label{Bx}
\end{align}
where $\mbf B$ is an $L^{\mrm s} \times N$ matrix with its $(l,n)$-th element given by $[\mbf B]_{l,n} = (\mbf b(\theta_l^{\mrm s},\phi_l^{\mrm s}) \otimes \mbf a_N(\tau_l^{\mrm s}))^H_n (\mbf y)_n$, and $(\cdot)_n$ denotes the $n$-th sub-vector of size $M \times 1$.
 Let $\mbf C \triangleq \sigma_w^{-2} (\sigma_w^2 \bsm{\Lambda}^{-1} + \mbb{E}[ \mbf A(0)^H \mbf A(0) ] )^{-1}$. Substituting these into \eqref{fD}, we have
\begin{align}
	f(\epsilon) 
	& = \mbf a_N(\epsilon)^T \mbf B^H \mbf C \mbf B \mbf a_N(\epsilon)^* \notag \\
	& = \sum_{n=0}^{N-1} \sum_{k=0}^{N-1} e^{-j(n-k)\epsilon} [\mbf B^H \mbf C \mbf B]_{n,k} = \sum_{m=-(N-1)}^{N-1} e^{-jm\epsilon} \eta_m, \label{fD_expanded}
\end{align}
where $\eta_m = \sum_{n-k=m} [\mbf B^H \mbf C \mbf B]_{n,k}$. Based on \eqref{fD_expanded}, $f(\epsilon)$ is written as
\begin{align}
	f(\epsilon) =  \operatorname{Re} \left\{ \bsm{\eta}_{\epsilon}^H \mbf a_N(\epsilon) \right\} + \text{const}, \label{fD_final}
\end{align}
where $\bsm{\eta}_{\epsilon} = 2 [\eta_0, \ldots, \eta_{N-1}]^T$. Following the procedure outlined in \eqref{eq:FFT_search}-\eqref{eq:bf_tau_approx}, we approximate the belief $b_{f_{\mbf Y}}(\epsilon)$ as a von Mises distribution $\mal{VM}(\epsilon; \hat{\epsilon}, \kappa_{\epsilon})$.

To obtain the belief for the path coefficients $\bsm{\beta}^{\mrm s}$, we marginalize out $\epsilon$ from the joint belief $b_{f_{\mbf Y}}(\bsm{\beta}^{\mrm s}, \epsilon)$ given in \eqref{b_fYt_beta_epsilon}. This requires evaluating the integral
\begin{align}
	b_{f_{\mbf Y}}(\bsm{\beta}^{\mrm s}) \propto \int \exp(g(\bsm{\beta}^{\mrm s}, \epsilon)) \mal{VM}(\epsilon; \hat{\epsilon}, \kappa_{\epsilon}) d\epsilon,
\end{align}
which is generally intractable. We approximate it by evaluating the integrand at the mean of $\epsilon$, namely $\epsilon = \hat{\epsilon}$. Under this approximation, the belief for $\bsm{\beta}^{\mrm s}$ becomes Gaussian:
\begin{align}
	b_{f_{\mbf Y}}(\bsm{\beta}^{\mrm s}) \approx \mathcal{CN}(\bsm{\beta}^{\mrm s}; \bsm{\mu}(\hat{\epsilon}), \bsm{\Sigma}(\hat{\epsilon})). \label{bf_beta_approx}
\end{align}
Then, we obtain the estimate $\hat{\bsm{\beta}}^{\mrm s} = \bsm{\mu}(\hat{\epsilon})$.

\subsection{Path Generation Mechanism}

A critical aspect of iterative algorithms is the initialization. We need to initialize $\{\tau_l^{\mrm{s}}, \theta_l^{\mrm{s}}, \phi_l^{\mrm{s}}\}_{l=1}^{L^{\mrm{s}}}$ when no informative prior is available. Moreover, each measurement is corrupted by unknown $\epsilon(t),\forall t$, which render standard initialization methods, such as DFT-codebook peak picking or orthogonal matching pursuit \cite{OMP}, ineffective, since they introduce different shifts at different time-slots. To address this issue, we propose a path generation mechanism. Rather than attempting to initialize all $L^{\mrm{s}}$ paths, this approach begins with an empty set of paths and detects and estimates the parameters of dominant paths in the algorithm iterations.

Without loss of generality, we suppose that $i-1$ paths have already been identified. To generate the $i$-th path, we first compute the residual signal across all $T$ time slots as
\begin{align}
	\mbf R_i(t) = \mbf Y^{\mrm s}(t) - \sum_{l=1}^{i-1} \hat{\beta}_l^{\mrm s}(t) \mbf a_N(\hat{\tau}_l^{\mrm s} + \hat{\epsilon}(t)) \mbf b(\hat{\theta}_l^{\mrm s}, \hat{\phi}_l^{\mrm s})^T, \label{residual_path_gen}
\end{align}
The residual signal $\mbf R_i(t)$ can be interpreted as a noisy observation of the contribution from the $i$-th path:
\begin{align}
	\mbf R_i(t) = \beta_i^{\mrm s}(t) \mbf a_N(\tau_i^{\mrm s} + \epsilon(t)) \mbf b(\theta_i^{\mrm s}, \phi_i^{\mrm s})^T + \widetilde{\mbf W}_i(t),
\end{align}
where $\widetilde{\mbf W}_i(t)$ encompasses residual interference from unmodeled paths and measurement noise.

We now sequentially estimate the parameters $\tau_i^{\mrm s}$, $\theta_i^{\mrm s}$, and $\phi_i^{\mrm s}$ for this new path. To estimate the delay $\tau_i^{\mrm s}$, we adopt a marginal likelihood approach. We treat the instantaneous path coefficients $\{\beta_i^{\mrm s}(t)\}_{t=1}^T$ and the angular parameters $\{\theta_i^{\mrm s}, \phi_i^{\mrm s}\}$ as the variables with non-informative priors and integrate them out. Specifically, define $\bsm{\alpha}_i = \beta_i^{\mrm s}(t) \mbf b(\theta_i^{\mrm s}, \phi_i^{\mrm s})$. The likelihood for a single measurement is
\begin{align}
	p(\mbf R_i(t) | \tau_i^{\mrm s}, \bsm{\alpha}_i) \propto \exp \left( -\frac{1}{\sigma_w^2} \| \mbf R_i(t) - \mbf a_N(\tau_i^{\mrm s} + \hat{\epsilon}(t)) \bsm{\alpha}_i^T \|_F^2 \right), \notag
\end{align}
By integrating out $\bsm{\alpha}_i$, we obtain the marginal likelihood for $\tau_i^{\mrm s}$ from all $T$ measurements:
\begin{align}
p(\{\mbf R_i(t)\}_{t=1}^T | \tau_i^{\mrm s}) \propto \exp \left( \frac{1}{\sigma_w^2 N} \sum_{t=1}^T \| \mbf a_N^H(\tau_i^{\mrm s} + \hat{\epsilon}(t)) \mbf R_i(t) \|_F^2 \right). \label{marginal_likelihood_tau}
\end{align}
Similar to \eqref{fD_expanded}-\eqref{fD_final}, the right-hand side of \eqref{marginal_likelihood_tau} can be manipulated in the form $\exp(\operatorname{Re}\{\bsm{\eta}^H \mbf a_N(\tau_i^{\mrm s})\})$. Then, we apply the FFT-based search and Newton method (in \eqref{eq:FFT_search}-\eqref{eq:bf_tau_approx}) to obtain the estimate $\hat{\tau}_i^{\mrm s}$.

Given $\hat{\tau}_i^{\mrm s}$, we next estimate the azimuth angle $\theta_i^{\mrm s}$. The marginal likelihood for $\theta_i^{\mrm s}$ is given by
\begin{align}
	p(\{\mbf R_i(t)\}_{t=1}^T | \theta_i^{\mrm s} ; \hat{\tau}_i^{\mrm s}) \propto \exp \left( \frac{1}{\sigma_w^2 N M_1} \sum_{t=1}^T \left\| \mbf a_{M_1}^H (\theta_i^{\mrm s}) \mbf Z_i(t) \right\|_2^2 \right), \label{p_R_theta}
\end{align}
where $\mbf Z_i(t)$ is an $M_1 \times M_2$ matrix obtained by reshaping the $M \times 1$ vector $\left( \sum_{n=0}^{N-1} e^{-j n (\hat{\tau}_i^{\mrm s} + \hat{\epsilon}(t))} \mbf r_{i,n}^T(t) \right)$ with $\mbf r_{i,n}$ being the $n$-th row of $\mbf R_i$. Then, we express the right-hand side in the form $\exp(\operatorname{Re}\{\bsm{\eta}^H \mbf a_{M_1}(\theta_i^{\mrm s})\})$ and obtain $\hat{\theta}_i^{\mrm s}$.

Given $\hat{\tau}_i^{\mrm s}$ and $\hat{\theta}_i^{\mrm s}$, we estimate the zenith angle $\phi_i^{\mrm s}$. The marginal likelihood for $\phi_i^{\mrm s}$ is obtained by
\begin{align}
	& p(\{\mbf R_i(t)\}_{t=1}^T | \phi_i^{\mrm s} ; \hat{\tau}_i^{\mrm s}, \hat{\theta}_i^{\mrm s}) \notag \\
	& \propto \exp \left( \frac{1}{ \sigma_w^2 N M_1 M_2} \sum_{t=1}^T | \mbf a_{M_2}^H(\phi_i^{\mrm s}) \mbf Z_i^T(t) \mbf a_{M_1}^*(\hat{\theta}_i^{\mrm s}) |^2 \right). \label{p_R_phi}
\end{align}
Similarly, we find the estimate $\hat{\phi}_i^{\mrm s}$. After obtaining $\{\hat{\tau}_i^{\mrm s}, \hat{\theta}_i^{\mrm s}, \hat{\phi}_i^{\mrm s}\}$, we add the $i$-th path to the current path set and update $\epsilon(t),\forall t$ and $\{\beta_l^{\mrm s}\}_{l=1}^i$ according to the method in Sec. V-B. Notably, by summing the projection energies of residual signals across all time slots (in \eqref{marginal_likelihood_tau}, \eqref{p_R_theta}, and \eqref{p_R_phi}), the algorithm accumulates geometric information from diverse user poses. This ensures that the common grid-level parameters $\{\tau_i^{\mrm s}, \theta_i^{\mrm s}, \phi_i^{\mrm s}\}$ are effectively extracted under the power spectral variation.

\subsection{Overall Algorithm}

\begin{algorithm}[h]
    \caption{Quasi-Static Parameter Estimation}
    \label{alg:slow_ckm}
    \begin{algorithmic}[1]
        \REQUIRE $\{\mbf Y^{\mrm s}(t)\}_{t=1}^T$, $L^{\mrm s}$, $I_{\text{sta}}$, path count $i=0$.
        \FOR{$iter = 1$ to $I_{\text{sta}}$}

		\STATE \textit{// Path generation:}
        \IF{stopping criterion is not met}
        \STATE $i \leftarrow i + 1$.
        \STATE Obtain initial $\hat{\tau}_i^{\mrm s}$ via \eqref{marginal_likelihood_tau}.
        \STATE Obtain initial $\hat{\theta}_i^{\mrm s}$ and $\hat{\phi}_i^{\mrm s}$ via \eqref{p_R_theta}-\eqref{p_R_phi}.
        \ENDIF
		
        \STATE \textit{// Update the delays and angles for paths $l=1,...,i$:}
        \STATE Compute residual $\hat{\mathbf{R}}_{t,l}$ via \eqref{residual}.
        \STATE Update  $b_{f_{\mbf Y}}(\tau_l^{\mrm s})$ via \eqref{eq:Sec4_g_tau}-\eqref{eq:Sec4_eta_g_tau} and \eqref{eq:FFT_search}-\eqref{eq:bf_tau_approx}.
        \STATE Update $b_{f_{\mbf Y}}(\theta_l^{\mrm s})$ and $b_{f_{\mbf Y}}(\phi_l^{\mrm s})$ via \eqref{g_theta_s}-\eqref{eta_theta_s}.

        \STATE \textit{// Update sync. errors and quasi-static coefficients:}
        \STATE Update $b_{f_{\mbf Y}}(\epsilon(t))$ via \eqref{fD_final}.
        \STATE Update $b_{f_{\mbf Y}}(\bsm{\beta}^{\mrm s}(t))$ via \eqref{bf_beta_approx}.
        \STATE Update $\hat{\rho}_l^{\mrm s} = \frac{1}{T} \sum_{t=1}^T |\hat{\beta}_l^{\mrm s}(t)|^2 + [\bsm{\Sigma}(\hat{\epsilon})]_{l,l}$.

        \ENDFOR

        \RETURN Quasi-static parameters $\{\hat{\tau}_l^{\mrm s}, \hat{\theta}_l^{\mrm s}, \hat{\phi}_l^{\mrm s}\}_{l=1}^{L^{\mrm{s}}}$.
    \end{algorithmic}
\end{algorithm}



We summarize the overall algorithm for the quasi-static parameter estimation in Algorithm \ref{alg:slow_ckm}. The iteration begins with a conditional path generation stage (in steps 3-7), where a new dominant path is identified and added to the path set if a stopping criterion is not met. This is followed by an iterative refinement stage (in steps 9-11 and 13-15), where the parameters of all currently generated paths, along with the synchronization errors and path coefficients, are updated. This process repeats until the maximum number of iterations is reached.
In practice, the noise variance $\sigma_w^2$ and the path power $\rho_l^{\mrm s}$  are typically unknown. We employ the expectation-maximization (EM) method \cite[Sec. IV-D]{Wenjun_CKM} to estimate them in each algorithm iteration.

The computational complexity of Algorithm 1 is dominated by the iterative refinement stage. In the $iter$-th iteration, consider that $i_{iter}$ paths have been identified. Updating the parameters for all $i_{iter}$ paths (in steps 9-11) requires matrix-vector multiplication and FFT-based searches, leading to a complexity of $O(i_{iter} T (NM + N \log N + M \log M))$. For updating $\epsilon(t)$ and $\bsm{\beta}^{\mrm{s}}(t)$ (in steps 13-14), the computation of $\eta_m$ in \eqref{fD_expanded} involves matrix multiplication and inversion, resulting in a complexity of $O(T(i_{iter}^3 + i_{iter}^2 N + i_{iter} (N^2+NM)))$. Therefore, the total complexity is $\sum_{iter=1}^{I_{\text{static}}} O(T(i_{iter}^3 + i_{iter}^2 N + i_{iter}(N^2 + NM + N \log N + M \log M)))$. Since the algorithm is performed with a relatively long period, this complexity is acceptable.


\section{Algorithm Design for Dynamic Parameter Estimation}


\subsection{Beliefs of Synchronization Error and Quasi-Static Coefficients}

In the stage of dynamic parameter estimation, we treat $\{\hat{\tau}_l^{\mrm{s}}, \hat{\theta}_l^{\mrm{s}}, \hat{\phi}_l^{\mrm{s}}\}_{l=1}^{L^{\mrm s}}$ as fixed. The unknowns for the quasi-static channel component are the instantaneous complex coefficients $\{\beta_l^{\mrm{s}}\}_{l=1}^{L^{\mrm{s}}}$ and the group delay offset $\epsilon$. The joint estimation of $\epsilon$ and the quasi-static coefficients $\{\beta_l^{\mrm{s}}\}$ is the computationally dominant part. We can reuse the vectorized model from Section V-B. The residual signal corresponding to the quasi-static paths is expressed as
\begin{align}
	\mbf y_{\mrm{res}} = \mbf y - \sum_{l=1}^{L^{\mrm{d}}} \hat{\beta}_l^{\mrm{d}} \mrm{vec}(\tilde{\mbf a}_N(\hat{\tau}_l^{\mrm{d}}) \mbf b(\hat{\theta}_l^{\mrm{d}}, \hat{\phi}_l^{\mrm{d}})^T). \label{residual_static}
\end{align}
In practice, due to the power spectral variation, the path power $\rho_l^{\mrm{s}}$ estimated from Algorithm 1 may not accurately reflect the instantaneous path power at the current snapshot. When computing $\ln b_{f_{\mbf Y}}(\bsm{\beta}^{\mrm s} , \epsilon)$, we omit the prior terms $\ln p(\bsm{\beta}^{\mrm s})$ and $\ln p(\epsilon)$, leading to
\begin{align}
	\ln b_{f_{\mbf Y}}(\bsm{\beta}^{\mrm s} , \epsilon) \propto \mathbb{E}_{\bsm{\xi} \setminus \{ \bsm{\beta}^{\mrm s}, \epsilon \}} \left[ \ln f_{\mbf Y}(\bsm{\xi}) \right]. \label{eq:b_fY_simplified}
\end{align}
Under this simplification, $g(\bsm{\beta}^{\mrm s},\epsilon)$ and $f(\epsilon)$ are expressed as
\begin{subequations}
	\label{eq:g_f_dynamic}
	\begin{align}
		g(\bsm{\beta}^{\mrm s},\epsilon) &= - (\bsm{\beta}^{\mrm s} - \bsm{\mu}(\epsilon))^H \bsm{\Sigma}^{-1} (\bsm{\beta}^{\mrm s} - \bsm{\mu}(\epsilon)), \\
		f(\epsilon) &= \sigma_w^{-2}\mbf y_{\mrm{res}}^{H} {\mbf A}(\epsilon) (\mbf A(0)^H \mbf A(0))^{-1} {\mbf A}(\epsilon)^H \mbf y_{\mrm{res}},
	\end{align}	
\end{subequations}
with $\bsm{\mu}(\epsilon) = (\mbf A(0)^H \mbf A(0))^{-1} {\mbf A}(\epsilon)^H \mbf y_{\mrm{res}}$ and $\bsm{\Sigma} = \sigma_w^2 (\mbf A(0)^H \mbf A(0))^{-1}$. The matrix inversion $(\mbf A(0)^H \mbf A(0))^{-1}$ is deterministic once $\{\hat{\tau}_l^{\mrm{s}}, \hat{\theta}_l^{\mrm{s}}, \hat{\phi}_l^{\mrm{s}}\}_{l=1}^{L^{\mrm s}}$ are given from the quasi-static parameter estimation of stage I. Therefore, it is computed only once during the algorithm iterations. Alternatively, $(\mbf A(0)^H \mbf A(0))^{-1}$ can be pre-computed and stored as part of the output in stage I, with a storage complexity $O((L^{\mrm s})^2)$.

\vspace{-0.2 cm}
\subsection{Beliefs of Dynamic Path Delays and Angles}

The estimation of $\{\tau_l^{\mrm{d}}, \theta_l^{\mrm{d}}, \phi_l^{\mrm{d}}\}$ follows a similar procedure to the esimation of $\{{\tau}_l^{\mrm{s}}, {\theta}_l^{\mrm{s}}, {\phi}_l^{\mrm{s}}\}$ in Sec. V-A, since their signal models are symmetric. A modification is the residual signal. For the $l$-th dynamic path, the residual signal is expressed as
\begin{align}
	\hat{\mathbf{R}}_{l}^{\mrm{d}} \triangleq &  \mathbf{Y}^{\mrm{d}} - \sum_{j=1}^{L^{\mrm{s}}} \hat{\beta}_j^{\mrm{s}} \tilde{\mathbf{a}}_N(\hat{\tau}_j^{\mrm{s}} + \hat{\epsilon})\, \mathbf{b}(\hat{\theta}_j^{\mrm{s}}, \hat{\phi}_j^{\mrm{s}})^T \notag \\
	&  - \sum_{k \neq l} \hat{\beta}_k^{\mrm{d}} \tilde{\mathbf{a}}_N(\hat{\tau}_k^{\mrm{d}})\, \mathbf{b}(\hat{\theta}_k^{\mrm{d}}, \hat{\phi}_k^{\mrm{d}})^T, \label{residual_dynamic}
\end{align}
Given \eqref{residual_dynamic}, we adopt the method in Sec. V-A to update the belief of $\{\tau_l^{\mrm{d}}, \theta_l^{\mrm{d}}, \phi_l^{\mrm{d}}\}$. Recall that $\epsilon$ is absorbed in $\tau_l^{\mrm{d}}$. For the initialization of $\{\tau_l^{\mrm{d}}, \theta_l^{\mrm{d}}, \phi_l^{\mrm{d}}\}$, we apply \eqref{marginal_likelihood_tau}-\eqref{p_R_phi}, where we replace $\{\hat{\tau}_l^{\mrm{s}}, \hat{\theta}_l^{\mrm{s}}, \hat{\phi}_l^{\mrm{s}}\}$ with $\{\tau_l^{\mrm{d}}, \theta_l^{\mrm{d}}, \phi_l^{\mrm{d}}\}$ and set $\hat{\epsilon}=0$.  Unlike stage I where the quasi-static paths are generated during iterations, all $L^{\mrm d}$ dynamic paths are generated in the initialization stage.

\subsection{Beliefs of Dynamic Path Coefficients}
For the dynamic path coefficients $\beta_l^{\mrm{d}},\forall l$, we adopt a path-wise estimation scheme. Specifically, we apply \eqref{b_fi_final} to obtain
\begin{align}
	b_{f_{\mbf Y^{\mrm{d}}}}(\beta_l^{\mrm{d}}) \propto g(\beta_l^{\mrm{d}}) p(\beta_l^{\mrm{d}}),
\end{align}
where $g(\beta_l^{\mrm{d}}) = \mathcal{CN}(\beta_l^{\mrm{d}}; \mu_{g,l}, v_{g,l})$, with the mean and variance given by
\begin{subequations}
	\label{eq:g_beta_dynamic}
	\begin{align}
	v_{g,l} &= \sigma_w^2 / \|\hat{\mbf a}_N({\tau}_l^{\mrm{d}}) \hat{\mbf b}({\theta}_l^{\mrm{d}}, {\phi}_l^{\mrm{d}})^T\|_F^2, \\
	\mu_{g,l} &= v_{g,l} \cdot \mrm{Tr}\left( (\hat{\mbf a}_N({\tau}_l^{\mrm{d}}) \hat{\mbf b}({\theta}_l^{\mrm{d}}, {\phi}_l^{\mrm{d}})^T)^H \hat{\mathbf{R}}_{l}^{\mrm{d}} \right) / \sigma_w^2.
\end{align}
\end{subequations}
Given $p(\beta_l^{\mrm{d}}) = (1 - \lambda_l^{\mrm{d,pri}})\, \delta(\beta_l^{\mrm{d}}) + \lambda_l^{\mrm{d,pri}}\, \mathcal{CN}(\beta_l^{\mrm{d}};0,v_l^{\mrm{d,pri}})$, the resulting belief $b(\beta_l^{\mrm{d}})$ is also a BG distribution:
\begin{align}
	b(\beta_l^{\mrm{d}}) = (1- \lambda_l^{\mrm{d,post}}) \delta(\beta_l^{\mrm{d}}) + \lambda_l^{\mrm{d,post}} \mathcal{CN}( \beta_l^{\mrm{d}} ; \mu_l^{\mrm{d}} , v_l^{\mrm{d}} ),
\end{align}
where
\begin{subequations}
	\label{eq:beta_dynamic}
	\begin{align}
	v_l^{\mrm{d}} &= \left( (v_l^{\mrm{d,pri}})^{-1} + (v_{g,l})^{-1} \right)^{-1}, \\
	\mu_l^{\mrm{d}} &= v_l^{\mrm{d}} \cdot (v_{g,l})^{-1} \mu_{g,l}, \\
	\lambda_l^{\mrm{d,post}} & = C_1/(C_0 + C_1),
\end{align}
\end{subequations}
with $C_0 = (1-\lambda_l^{\mrm{d,pri}}) \mathcal{CN}(0; \mu_{g,l}, v_{g,l})$ and $C_1 = \lambda_l^{\mrm{d,pri}} \mathcal{CN}(0; \mu_{g,l}, v_l^{\mrm{d,pri}} + v_{g,l})$. The estimate of $\beta_l^{\mrm{d}}$ is then $ \hat{\beta}_l^{\mrm{d}} = \lambda_l^{\mrm{d,post}} \mu_l^{\mrm{d}}$.

\subsection{Overall Algorithm}

\begin{algorithm}[h]
	\caption{Dynamic Parameter Estimation}
	\label{alg:fast_ckm}
	\begin{algorithmic}[1]
		\REQUIRE  $\mbf Y^{\mrm d}$,  $\{\hat{\tau}_l^{\mrm s}, \hat{\theta}_l^{\mrm s}, \hat{\phi}_l^{\mrm s}\}_{l=1}^{L^{\mrm s}}$,  $L^{\mrm d}$, $I_{\text{dyn}}$.

		\STATE Initialize $\bsm{\beta}^{\mrm{s}}$ and $\epsilon$ via \eqref{residual_static} and \eqref{eq:g_f_dynamic} with $\hat{\beta}_l^{\mrm{d}} = 0, \forall l$.
		\STATE Initialize $\{\tau_l^{\mrm{d}}, \theta_l^{\mrm{d}}, \phi_l^{\mrm{d}}\}_{l=1}^{L^{\mrm{d}}}$ based on \eqref{marginal_likelihood_tau}-\eqref{p_R_phi}.
		
		\FOR{$iter = 1$ to $I_{\text{dyn}}$}
		\STATE \textit{// Update quasi-static coefficients and sync. error:}
		\STATE Update $b_{f_{\mbf Y}}(\epsilon)$ and $b_{f_{\mbf Y}}(\bsm{\beta}^{\mrm s})$ via \eqref{residual_static}-\eqref{eq:g_f_dynamic}.
		
		\STATE \textit{// Update dynamic parameters:}
		\STATE Update residual signal $\hat{\mathbf{R}}_{l}^{\mrm{d}}$ via \eqref{residual_dynamic}.
		\STATE Update beliefs of $\tau_l^{\mrm d}, \theta_l^{\mrm d}, \phi_l^{\mrm d}$ according to Sec. VI-B.
		\STATE Update belief of $\beta_l^{\mrm d}$ via \eqref{eq:g_beta_dynamic}-\eqref{eq:beta_dynamic}.
		\ENDFOR
		
		\RETURN $\{\hat{\beta}_l^{\mrm s}\}, \hat{\epsilon}, \{\hat{\beta}_l^{\mrm d}, \hat{\tau}_l^{\mrm d}, \hat{\theta}_l^{\mrm d}, \hat{\phi}_l^{\mrm d}\}$.
	\end{algorithmic}
\end{algorithm}



The overall algorithm for the dynamic parameter estimation is summarized in Algorithm \ref{alg:fast_ckm}. The algorithm begins with an initialization phase (steps 1-2), where the the initial values of $\beta_l^{\mrm{s}}$ and $\epsilon$ are estimated given $\{\hat{\tau}_l^{\mrm s}, \hat{\theta}_l^{\mrm s}, \hat{\phi}_l^{\mrm s}\}_{l=1}^{L^{\mrm s}}$, and the dynamic paths are detected. This is followed by an iterative refinement phase (steps 4-9), where the beliefs of all unknown parameters—including $\beta_l^{\mrm{s}}$, $\epsilon$, and $\{\beta_l^{\mrm{d}}, \tau_l^{\mrm{d}}, \theta_l^{\mrm{d}}, \phi_l^{\mrm{d}}\}$—are iteratively updated until convergence. In each iteration, we adopt the EM method \cite[Sec. IV-D]{Wenjun_CKM} to estimate $\sigma_w^2$, $\{\lambda_l^{\mrm{d,pri}}\}$ and $\{v_l^{\mrm{d,pri}}\}$.

A pivotal application of Algorithm 2 is channel estimation. By using a subset of subcarriers (i.e., $P < N$), Algorithm \ref{alg:fast_ckm} extracts the refined quasi-static parameters $\{\hat{\beta}_l^{\mrm s}, \hat{\epsilon}\}$ and the dynamic path parameters $\{\hat{\beta}_l^{\mrm d}, \hat{\tau}_l^{\mrm d}, \hat{\theta}_l^{\mrm d}, \hat{\phi}_l^{\mrm d}\}$. These estimates can reconstruct the channel response across the $N$ subcarriers as
$	\hat{\mbf{H}}^{\mrm{BB}} =  \sum_{l=1}^{L^{\mrm{s}}} \hat{\beta}_l^{\mrm{s}} \mbf a_N(\hat{\tau}_l^{\mrm{s}} + \hat{\epsilon}) \mbf b(\hat{\theta}_l^{\mrm{s}}, \hat{\phi}_l^{\mrm{s}})^T 
	 + \sum_{l=1}^{L^{\mrm{d}}} \hat{\beta}_{l}^{\mrm{d}} \mbf a_N(\hat{\tau}_{l}^{\mrm{d}}) \mbf b(\hat{\theta}_{l}^{\mrm{d}}, \hat{\phi}_{l}^{\mrm{d}})^T. 
$

The computational complexity of Algorithm \ref{alg:fast_ckm} is analyzed as follows. The initialization of dynamic paths (step 2) has a complexity of $\mathcal{O}(L^{\mrm d}(PM + P \log P + M \log M))$. Updating $\epsilon$ and $\{\beta_l^{\mrm s}\}$ (step 5) requires $\mathcal{O}((L^{\mrm s})^3 + (L^{\mrm s})^2 P + L^{\mrm s}(PM+P^2))$. Updating the parameters for all $L^{\mrm d}$ dynamic paths (steps 7-9) has a complexity of $\mathcal{O}(L^{\mrm d}(PM + P \log P + M \log M))$. Therefore, the overall complexity is $\mathcal{O}((L^{\mrm s})^3 + I_{\mrm{dyn}} (L^{\mrm s})^2 P + I_{\mrm{dyn}} L^{\mrm s}(PM+P^2) + I_{\mrm{dyn}} L^{\mrm d}(PM + P \log P + M \log M))$. Since $L^{\mrm d}$ is typically smaller than $L^{\mrm s}$, the dominant terms are $\mathcal{O}((L^{\mrm s})^3 + I_{\mrm{dyn}} (L^{\mrm s})^2 P + I_{\mrm{dyn}} L^{\mrm s}(PM+P^2))$. The cubic term $\mathcal{O}((L^{\mrm s})^3)$ arises from the inversion $(\mbf A(0)^H \mbf A(0))^{-1}$. This matrix inversion can be pre-computed (in stage I) to save computation, with a storage of $\mathcal{O}((L^{\mrm s})^2)$. In simulations, as shown in Sec. VII, the algorithm converges within a few iterations (e.g., $5$ iterations) and operates with $P \ll N$, which further reduces the computational complexity.

\vspace{-0.2 cm}
\section{Numerical Results}

\subsection{Simulation in the LoS Scenario of 28 GHz} \label{sec:simu_LOS}

In this subsection, we conduct numerical simulations to evaluate the performance of the proposed CKM construction algorithm. The system parameters are set as follows: the carrier frequency is $28$ GHz, the number of OFDM subcarriers is $N = 192$ with a subcarrier spacing of $30$ kHz, and the BS antenna array size is $M_1 \times M_2 = 4 \times 8$. The antenna radiation patterns for both the BS and the user follow the 3GPP-3D model specified in TR 36.873 \cite{TR36873}.

For the quasi-static channel component, the BS coverage area is divided into $1 \times 1 \text{ m}^2$ square grids. Within each grid, channel measurements are generated along a trajectory following a random walk process, with the user antenna orientation aligned with the direction of motion at each sampling point. We adopt the UMa LoS channel model from 3GPP TR 38.901 \cite{TR38901}, which consists of one LoS path and 200 NLoS paths grouped into 10 clusters of 20 sub-paths each. The path delays and angles are generated according to \cite[Table 7.5-6]{TR38901}, and the spatial consistency procedure from \cite[Sec. 7.6.3.2 Procedure B]{TR38901} is applied to model the variation of channel parameters across different user poses. For the dynamic channel component, we consider $2$ dynamic scatterers, each with an activity probability of $0.5$. When active, a scatterer introduces a cluster of $10 - 20$ sub-paths with a $15$-degree angular spread and a random cluster delay uniformly chosen from $[0, 1]~\mu$s. The synchronization error for each measurement is drawn uniformly from $[0, 1]~\mu$s.

In stage I, we consider that the BS uses a dataset containing only quasi-static channel components at a relatively high SNR of $25$ dB. In practice, such a dataset can be obtained by excluding measurements with significant Doppler spreads induced by dynamic scatterers.
In stage II, the BS uses measurements that include both static and dynamic components, with a static-to-dynamic power ratio of $10:1$ and a SNR of $5$ dB. The estimation of quasi-static parameters use $18$ OFDM symbols within a grid; the estimation of dynamic parameters uses a single OFDM symbol and $P \le N$ pilot subcarriers.

To evaluate the method of quasi-static parameter estimation, we compare it against the following benchmarks:
\begin{itemize}
	\item \textbf{Perfect Sync.}: Assume known $\epsilon(t)$.
	\item \textbf{Without sync. error est.}: Ignore the estimation of  $\epsilon(t)$, i.e., set $\hat{\epsilon}(t)=0$.
	\item \textbf{Separate est.}: Estimate $\epsilon(t)$ and $\bsm{\beta}^{\mrm{s}}$ separately, replacing the joint estimation method in Sec. V-B.
	\item \textbf{OMP Init.}: Adopt the OMP algorithm \cite{OMP} for initialization, replacing the path generation in Sec. V-C.
\end{itemize}

\begin{figure}[h]
	\centering
	\includegraphics[width=0.45\textwidth]{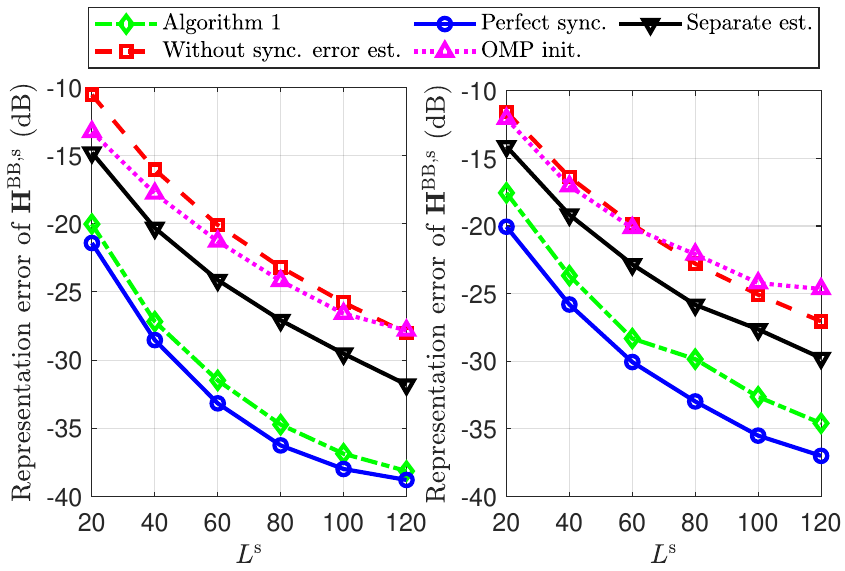}
	\caption{Representation error of $\mbf H^{\mrm{BB},\mrm{s}}$ versus $L^{\mrm{s}}$ in stage I (left subplot) and II (right subplot).}
	\label{fig:MSE_vs_L_dyn_LOS}
	\vspace{-0.2 cm}
\end{figure}

Fig. \ref{fig:MSE_vs_L_dyn_LOS} shows the representation error of $\mbf H^{\mrm{BB},\mrm{s}}$ versus $L^{\mrm{s}}$ in stage I (left subplot) and stage II (right subplot). We apply Algorithm 1 to obtain $\{\hat{\tau}_l^{\mrm s}, \hat{\theta}_l^{\mrm s}, \hat{\phi}_l^{\mrm s}\}_{l=1}^{L^{\mrm{s}}}$ and compute the representation error via \eqref{eq:representation_error}. In the left sub-plot, the performance of all algorithms improves as $L^{\mrm{s}}$ increases. For Algorithm 1, increasing $L^{\mrm{s}}$ from $40$ to $80$ improves the representation performance from $-27.5$ dB to $-35$ dB. At $L^{\mrm{s}}=80$, Algorithm 1 outperforms OMP init., Separate est., and Without sync. error. est. by $10.5$ dB, $8$ dB, and $11.5$ dB, respectively. The performance gap between Algorithm 1 and the Perfect Sync. benchmark remains within $1.5$ dB. In the right sub-plot, the performance gap among Algorithm 1 and benchmarks is similar. Besides, the NMSE values of all schemes in stage II are higher than those in stage I. This is because the channel data in stage II correspond to different user poses compared to those in stage I, leading to variations of quasi-static channel parameters.

Next, we evaluate the performance of the dynamic parameter estimation. We consider the following benchmarks:
\begin{itemize}
	\item \textbf{Without sync. error est.}: Ignore the estimation of $\epsilon(t)$, i.e., set $\hat{\epsilon}(t)=0$.
	\item \textbf{Without dynamic param. est.}: Ignore the estimation of $\mbf H^{\mrm{BB},\mrm{d}}$, i.e., set $\hat{\mbf H}^{\mrm{BB},\mrm{d}}=0$. 
	\item \textbf{Without quasi-static prior}: Do not exploit the estimation of quasi-static parameters (from stage I) as priors.
	\item \textbf{Ideal quasi-static prior}: Use the true $201$ path parameters from the UMa LoS model and known $\epsilon(t)$.
\end{itemize}

\vspace{-0.2 cm}
\begin{figure}[h]
	\centering
	\includegraphics[width=0.38\textwidth]{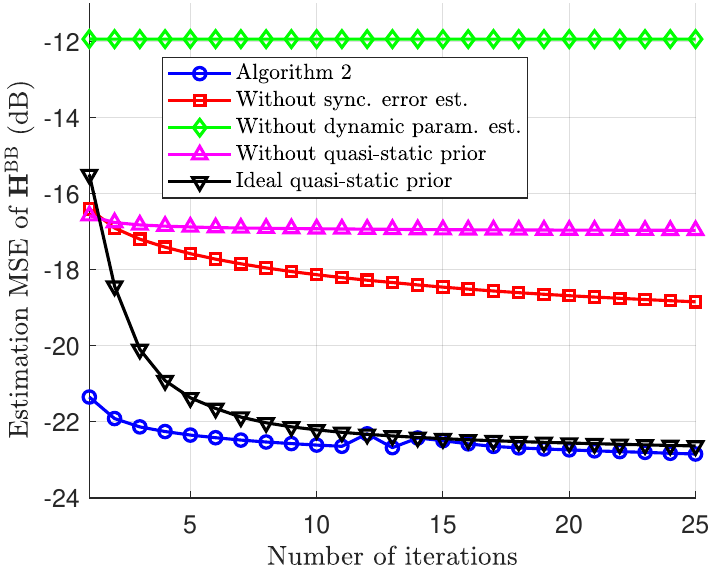}
	\caption{MSE of $\mbf H^{\mrm{BB}}$ versus the number of iterations in stage II.}
	\label{fig:MSE_vs_L_dyn_LOS2}
\end{figure}

Fig. \ref{fig:MSE_vs_L_dyn_LOS2} shows the estimation MSE of $\mbf H^{\mrm{BB}}$ versus the number of iterations in stage II, where we set $L^{\mrm{s}}=60$, $L^{\mrm{d}}=10$ and $P=N$. Algorithm II converges rapidly within $5$ iterations and achieves $>4.5$ dB performance improvement compared to the benchmarks. Particularly, Algorithm II slightly outperforms the ``Ideal quasi-static prior'' benchmark by $0.2$ dB. This slight superiority arises because Algorithm II reconstructs $\mbf H^{\mrm{BB}}$ using $L^{\mrm{s}}=60$ and $L^{\mrm{d}}=10$, whereas the ideal scheme utilizes the true $201$ quasi-static paths alongside the dynamic components. Although the reduced path set introduces a representation error of approximately $-28.5$ dB (as illustrated in the right subplot of Fig. \ref{fig:MSE_vs_L_dyn_LOS}), focusing on dominant paths mitigates overfitting under limited pilot overhead.
\begin{figure}[h]
	\centering
	\includegraphics[width=0.38\textwidth]{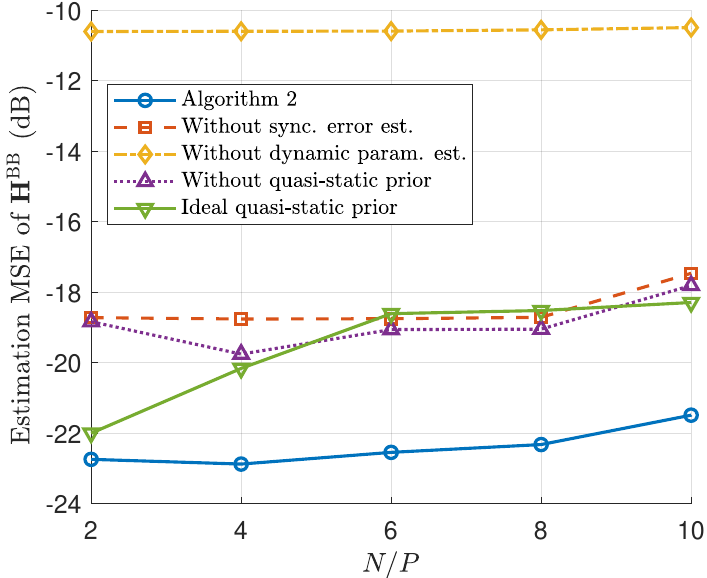}
	\caption{MSE of $\mbf H^{\mrm{BB}}$ versus $N/P$ in stage II.}
	\label{fig:LoS_Scenario_Full_MSE}
	\vspace{-0.2 cm}
\end{figure}

Fig. \ref{fig:LoS_Scenario_Full_MSE} shows the MSE performance versus the $N/P$ ratio in stage II, where we set $L^{\mrm{s}} = 40$ and $L^{\mrm{d}} = 40$. When $\mbf H^{\mrm{BB},\mrm{d}}$ is not estimated, the MSE remains constant at $-10.5$ dB. The other benchmarks and Algorithm II exhibit performance degradation as $N/P$ increases. For algorithm II, the MSE degrades from $-22.7$ dB to $-21.5$ dB as $N/P$ increases from $2$ to $10$, with a performance loss of $1.2$ dB. The performance gap between Algorithm II and the ``Ideal quasi-static prior'' benchmark widens as $N/P$ increases. These observations indicate that Algorithm II is robust to limited observations.

\vspace{-0.2 cm}
\subsection{Simulation in the NLoS Scenario of 6.5 GHz}

In this subsection, we consider the UMa NLoS scenario defined in 3GPP TR 38.901, with a carrier frequency of 6.5 GHz \cite{TR38901}. Compared to the UMa LoS case, the NLoS environment features a richer scattering environment with $20$ clusters, each containing $20$ sub-paths, for a total of $400$ paths. Other system parameters remain the same as in Sec. \ref{sec:simu_LOS}.

\begin{figure}[h]
	\centering
	\includegraphics[width=0.45\textwidth]{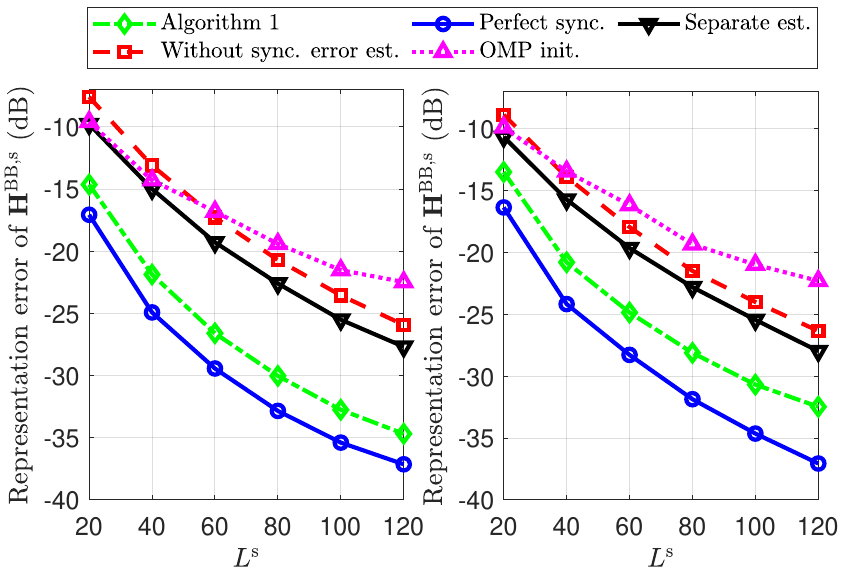}
	\caption{Representation error versus $L^{\mrm{s}}$ in stage I (left subplot) and II (right subplot).}
	\label{fig:MSE_vs_L_dyn_NLOS}
		\vspace{-0.2 cm}
\end{figure}

Fig. \ref{fig:MSE_vs_L_dyn_NLOS} shows the representation error of $\mbf H^{\mrm{BB},\mrm{s}}$ versus $L^{\mrm{s}}$ in stage I (left subplot) and stage II (right subplot). The performance gap between Algorithm I and the benchmarks follows a similar trend to the LoS scenario. However, the NMSE values for all schemes are higher in the NLoS case. At $L^{\mrm{s}}=80$, the NMSE of Algorithm I is $-30$ dB in the NLoS case, which is $5$ dB higher than that in the LoS case. This performance difference is because the channel energy is more dispersed across $400$ multipath components compared to the $201$ paths in the LoS scenario. Consequently, a fixed number of representative paths $L^{\mrm{s}}$ captures a smaller proportion of the total channel energy, resulting in a higher representation error.

\begin{figure}[h]
	\centering
	\includegraphics[width=0.38\textwidth]{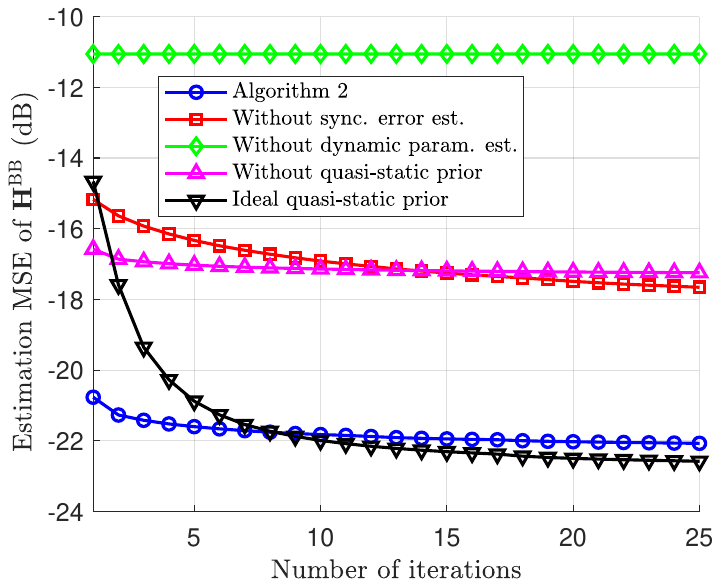}
	\caption{MSE of $\mbf H^{\mrm{BB}}$ versus the number of iterations in stage II.}
	\label{fig:MSE_vs_L_dyn_NLOS2}
	\vspace{-0.2 cm}
\end{figure}

Fig. \ref{fig:MSE_vs_L_dyn_NLOS2} shows the estimation MSE of $\mbf H^{\mrm{BB}}$ versus the number of iterations in stage II, where we set $L^{\mrm{s}}=60$, $L^{\mrm{d}}=10$ and $P=N$. Algorithm II converges within $5$ iterations and provides a performance gain of over $4.5$ dB compared to the  benchmarks. Besides, Algorithm II shows a slight performance degradation of $0.5$ dB compared to the Ideal quasi-static prior benchmark. We conjecture that this minor degradation is due to a higher representation error in the NLoS scenario.

\begin{figure}[h]
	\centering
	\includegraphics[width=0.38\textwidth]{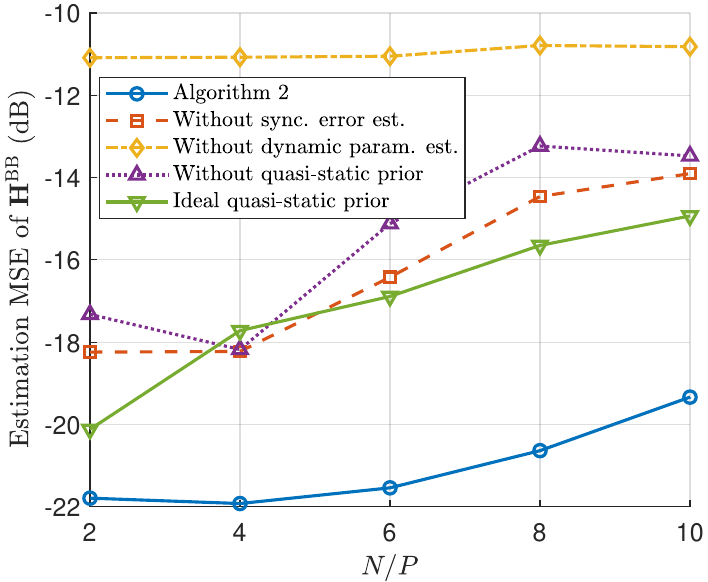}
	\caption{MSE of $\mbf H^{\mrm{BB}}$ versus $P/N$ in stage II.}
	\label{fig:NLoS_Scenario_Full_MSE}
	\vspace{-0.2 cm}
\end{figure}

Fig. \ref{fig:NLoS_Scenario_Full_MSE} shows the MSE performance versus the $N/P$ ratio in stage II, where we set $L^{\mrm{s}} = 60$ and $L^{\mrm{d}} = 10$. When $N/P$ increases from $2$ to $10$, the MSE of Algorithm II, the Ideal quasi-static prior, and the "Without quasi-static prior" schemes degrade by $2.5$ dB, $5$ dB, and $4.2$ dB, respectively, where Algorithm II exhibits the smallest performance loss. Compared to the LoS scenario in Fig. \ref{fig:LoS_Scenario_Full_MSE}, these performance degradations are higher, which indicates that the dynamic parameter estimation is more sensitive to pilot overhead when the propagation environment is more rich-scattering.

\vspace{-0.2 cm}

\section{Conclusion}

In this paper, we proposed a two-stage method for the dynamic CKM construction in MIMO-OFDM systems. We extracted grid-level quasi-static channel parameters from historical data within a Bayesian inference framework. These parameters were then used as strong informative priors for the second stage, which estimated dynamic channel parameters from limited measurements with low complexity. Extensive simulations show that the proposed scheme outperforms the benchmarks and enables low-overhead channel estimation.

\vspace{-0.2 cm}

\bibliographystyle{IEEEtran}
\bibliography{TurboMP}

\end{document}